\documentclass{aa}  

\usepackage{graphicx}
\usepackage{natbib}
\usepackage{float}
\usepackage{url}
\usepackage{txfonts}
\begin{document}

\title{Galactic or extragalactic chemical tagging for NGC3201?}

   \subtitle{Discovery of an anomalous CN-CH relation\thanks{Observations done under programme 60.A-9501(B) at NTT/ESO, La Silla; and archival data from project 60.A-9700(D).} \fnmsep
\thanks{Table 2 is also available in electronic form
at the CDS via anonymous ftp to cdsarc.u-strasbg.fr (130.79.128.5)
or via http://cdsweb.u-strasbg.fr/cgi-bin/qcat?J/A+A/}}

   \author{B. Dias
          \inst{1}
          \and
          I. Araya\inst{2,3}
          \and
          J.P. Nogueira-Cavalcante\inst{4,5}
        \and
                L. Saker\inst{6}
          \and
          A. Shokry\inst{1,7}
          }

   \institute{European Southern Observatory, Alonso de C\'ordova 3107, Santiago, Chile\\
              \email{bdias@eso.org}
         \and
             Instituto de F\'{\i}sica y Astronom\'{\i}a, Facultad de Ciencias, Universidad de Valpara\'{\i}so, Casilla 5030, Valpara\'{\i}so, Chile
             \and
             {N\'ucleo de Matem\'aticas, F\'isica y Estad\'istica, Facultad de Ciencias, Universidad Mayor, Chile}
             \and 
             Observat\'orio do Valongo, Universidade Federal do Rio de Janeiro, Ladeira Pedro Ant\^onio, 43, Sa\'ude 20080-090, Rio de Janeiro, Brazil
             \and
             Observat\'orio Nacional, Rua Gal. Jos\'e Cristino 77, S\~ao Crist\'ov\~ao 20921-400 Rio de Janeiro RJ, Brazil
             \and 
             Observatorio Astron\'omico, Universidad Nacional de C\'ordoba, Laprida 854, C\'ordoba, CP 5000, Argentina
             \and 
             National Research Institute of Astronomy and Geophysics (NRIAG), 11421 Helwan, Cairo, Egypt
             }

   \date{Received ; accepted }

 
  \abstract
   {The origin of the globular cluster (GC) \object{NGC~3201} is under
     debate. Its retrograde orbit points to an extragalactic
     origin, but no further chemical evidence supports this idea.
Light-element chemical abundances are useful
     to tag GCs and can be used to shed light on this discussion.
} 
   {Recently it was shown that the CN and CH indices are useful to
     identify GCs that are anomalous to those typically found in the Milky Way. 
     A possible origin of anomalous clusters is the merger of two GCs and/or 
     the nucleus of a dwarf galaxy.
 We aim to derive CN
     and CH band strengths for red giant stars in NGC3201 and compare these with
     photometric indices and high-resolution spectroscopy and discuss in the 
     context of GC chemical tagging.
     } 
   {We measure molecular band indices of S(3839) and G4300 for CN and
     CH, respectively from low-resolution spectra of red giant stars. Gravity and
     temperature effects are removed. 
     Photometric indices are used to indicate further chemical information on C+N+O or s-process
     element abundances that are not derived from low-resolution spectra.
     } 
   { We found three groups in the CN-CH distribution. A main sequence ({\it S1}), a secondary less-populated
   sequence ({\it S2}), and a group of peculiar ({\it pec}) CN-weak and CH-weak stars,
   one of which was previously known. 
   The three groups seem to have different C+N+O and/or s-process element abundances, to be confirmed by high-resolution spectroscopy.
   These are typical characteristics of anomalous GCs.
   The CN distribution of NGC~3201 is quadrimodal, which is more common in anomalous clusters.
   However, NGC~3201 does not belong to the trend of anomalous GCs in the mass-size relation. 
} 
{The globular cluster NGC~3201 shows signs that it can be chemically tagged as anomalous: it has an unusual CN-CH relation, indications that {\it pec}-{\it S1}-{\it S2} is an increasing sequence of C+N+O or s-process element abundances, and a multi-modal CN distribution that seems to correlate with s-process element abundances. The non-anomalous characteristics are that it has a debatable Fe-spread and it does not follow the trend of mass size of all anomalous clusters. 
Three scenarios are postulated here: (i) if the sequence {\it pec-S1-S2} has increasing C+N+O and s-process element abundances, NGC~3201 would be the first anomalous GC outside of the mass-size relation; (ii) if the abundances are almost constant, NGC~3201 would be the first non-anomalous GC with multiple CN-CH anti-correlation groups; or (iii) it would be the first anomalous GC without variations in C+N+O and s-process element abundances. In all cases, the definition of anomalous clusters and the scenario in which they have an extragalactic origin must be revised.
}


   \maketitle
%

\section{Introduction}

The paradigm that defines globular clusters (GCs) has
slowly changed since the first studies pointing to multiple populations
such as \cite{cottrell+81} and \cite{norris+81N6752}. A substantial amount of work has been
done based on high-resolution spectroscopic and photometric 
observations during the last two
decades. It is well known that all globular clusters present a
star-to-star variation of light-element abundances, such as the Na-O
anti-correlation (e.g. \citealp{gratton+04,gratton+12}) or the C-N 
anti-correlation 
(e.g. 
\citealp{cohen+02,briley+04,dacosta+04,kayser+08,pancino+10,smolinski+11}). The most 
plausible scenario to explain the anti-correlations is self-enrichment
\citep[e.g.][]{prantzos+06}, where the nature of the polluters is the current
open question of this field. Asymptotic giant branch (AGB) stellar ejecta \citep[e.g.][]
{dercole+08}, fast-rotating massive stars
\citep[e.g.][]{decressin+07}, and massive binaries \citep[e.g.][]
{demink+09} are among the possible candidates. In all cases, 
nucleosynthesis and stellar evolution have been discarded because 
un-evolved main sequence stars also present the anti-correlations 
above \citep[e.g.][]{briley+94,kayser+08,pancino+10,milone+13}. Therefore the origin of 
the primordial chemical anomalies in GCs is environmental and still remains
unknown \citep[see e.g.][]{renzini+15}. For an updated review on the recent scenarios 
and comparisons with observational constraints we refer to \cite{bastian+18}.

A few GCs also present a star-to-star spread in metallicity, C+N+O, and s-process element abundances,
where each group has a spread in p-capture element abundances, such as Na-O and C-N anti-correlation, 
for all cases when abundances are available \citep[more details in][]{dacosta15,marino+15}. They are called
anomalous GCs and are usually associated to
a peculiar formation and evolution history, possibly
originating in dwarf galaxies that were 
captured by the Milky Way \citep[e.g.][]{dacosta15}. 
They are important targets
to be analysed also in terms of light-element anti-correlations described 
above to characterise these clusters as typical globular clusters
or anomalous. One of these targets is M~22 \citep{dacosta+09,marino+09}, although \cite{mucciarelli+15b} 
 argued against a [Fe/H] spread for M~22. 
\cite{marino+11} showed that the CN index S(3839)  
traces [N/Fe] and the CH index G4300 traces [C/Fe]; they split the stars 
into two groups in terms of {\it s-process} element abundances. The s-poor
group has a bimodal 
C-N anti-correlation, as expected from 
CNO-cycle enrichment, and the s-rich group also has an anti-correlation 
but the sample is small, therefore it is not possible to say whether the distribution is 
bifurcated or not. 
Consequently M~22 presents two groups with C-N anti-correlation
and not a broad correlation as \cite{norris+83} concluded \citep[see also][]{lim+17}. 

The anomalous cluster \object{NGC~1851} is also potentially a result of the merger of two globular
clusters \citep[][]{carretta+10n1851}. As in the case of \object{M~22}, no 
CN-CH anti-correlation was found at first sight by \cite{lardo+12}. However,
a more detailed analysis revealed a strong anti-correlation of [C/Fe] and [N/Fe]
by Lardo et al. Another interesting result is that NGC~1851 does not present
a bimodal distribution of CN no [N/Fe], but does present a quadrimodal CN distribution 
as shown by \cite{campbell+12}. A closer look at the bimodal CN distribution 
of M~22 also reveals a quadrimodal distribution (see Sect. \ref{sec:cndist}.)
A difference between M~22
and NGC~1851 is that the latter presents a diffuse stellar halo that could be
a remnant of its host dwarf galaxy after being captured by the Milky Way
\citep{olszewski+09,bekki+12}. However, if M~22 had a stellar halo, it was
possibly stripped off because the cluster is closer to the Galactic centre
\citep{dacosta15}.
In conclusion, the formation scenario for
both M~22 and NGC~1851 points to a
merger of globular clusters, however the
details may not be entirely the same.
A few other anomalous clusters can be added to this group: 
\object{NGC~5286}, \object{M~2}, 
\object{NGC~5824}, \object{M~19}, \object{M~54}, and \object{M~75} \citep{dacosta15,marino+15}.

Another intriguing cluster is NGC~3201.
Although \cite{simmerer+13} showed that this object is among the clusters
with an intrinsic [Fe/H] spread, other works have strongly suggested otherwise
\citep{covey+03,munoz+13,mucciarelli+15a}. Nevertheless,
NGC~3201 has a retrograde orbit \citep{gonzalez+98}, which indicates a 
contentious history for this GC, with a possible extragalactic origin, 
similar to the case of $\omega$Cen \citep{bekki+03}.
It is a good candidate to present a peculiar CN-CH relation as in the 
case of M~22.
\cite{smith+82} found a bimodal distribution of CN for NGC~3201 red giant branch (RGB)
stars, but not as marked as for other typical GCs, such as \object{NGC~6752} 
and M~4. They also found a mild CN-CH anti-correlation but with one 
star that was CN-weak and CH-weak, in agreement with \cite{dacosta+81}.
A new analysis of CN and CH for NGC~3201 in comparison with the latest 
high-resolution spectroscopy, photometric 
data, and CN and CH of typical and anomalous
GCs is needed.
In this work we analyse CN and CH 
indices of RGB stars from NGC~3201, and discuss
the origin of this GC as done for other anomalous
GCs.

The paper is organised as follows.
In Sect. 2 we describe the photometric and spectroscopic observations.
The spectroscopic analysis and calibrations are done in Sect. 3. The multiple 
stellar generations of NGC~3201 are discussed in Sect. 4 and compared 
to other anomalous clusters.
Finally in Sect. 5 we discuss the origin of NGC~3201. 
Summary and conclusions are given in Sect. 6.

%
\section{The data}
\subsection{Target selection from pre-images}

The observations were carried out using the European Southern Observatory (ESO) Faint Object Spectrograph and Camera (v.2), EFOSC2 \citep{snodgrass+08}, mounted on the 3.6 m New Technology Telescope (NTT) at the ESO-La Silla Observatory, Chile.
We used archival images in B and V filters centred at NGC3201 to select RGB stars and prepare the masks in this region. To obtain more RGB stars we observed additional pointings to the north and to the south of the cluster; we observed also in B and V filters and kept half of the cluster in the field of view of 4$\arcmin$x4$\arcmin$ (see Table \ref{log}). Data reduction was done using ESO pipeline {\it esorex.}\footnote{\url{http://www.eso.org/sci/software/cpl/esorex.html}.}

\begin{figure}[!htb]
\centering
\includegraphics[height=\columnwidth, angle=-90]{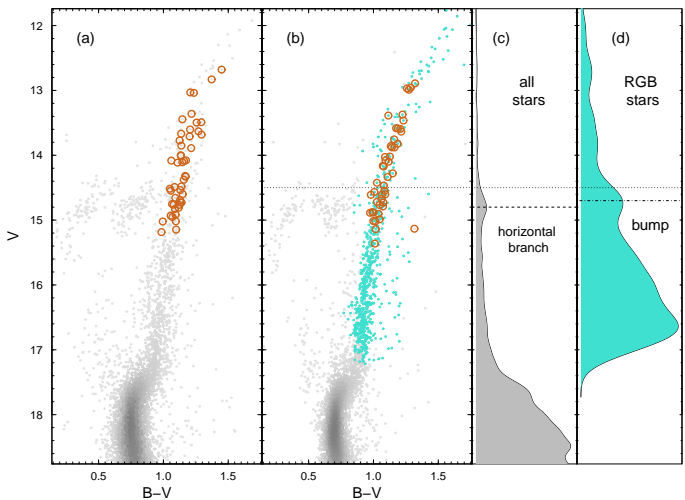}
\caption{Colour-magnitude diagram (CMD) of NGC~3201 and selected stars for spectroscopic observations. 
{\bf (a)} Photometry from pre-images as described in the text. No calibration or de-reddening process was done, except by
a zero point offset to match the calibrated CMD from panel (b). Targets for spectroscopic observations were selected from this CMD and are shown as orange circles. 
{\bf (b)} Same as (a) but using photometry from \cite{kravtsov+09} that is calibrated and corrected by differential reddening. The selected targets are identified and reveal an outlier from the RGB region that should be excluded from our analysis. Moreover, RGB stars are colour-coded. 
{\bf (c)} Luminosity function of all stars from (b) indicating the over density of the horizontal branch. 
{\bf (d)} Luminosity function only of RGB stars from (b) indicating the over density of the RGB bump.}
\label{cmd-mask1}
\end{figure}

To select RGB stars we carried out point spread function (PSF) photometry on B and V images using default procedures with DAOPHOT at image reduction and analysis facility software (IRAF) \citep{stetson+87}. From the colour-magnitude diagram (CMD, see Fig. \ref{cmd-mask1}) we selected all RGB stars brighter than V $<$ 15 and identified them in the pre-image. We note that the broad RGB is due to differential reddening \citep{kravtsov+09,vonbraun+01}. We selected the best targets that would provide spectra with features of CN, CH, and Fe on the detector without overlapping with neighbouring stars. In total we selected 
46 RGB stars in NGC3201. From Fig. \ref{cmd-mask1}b one star is located at the AGB phase at about V$\sim$13.4~mag and a further three brighter stars at about V$\sim$13.0~mag cannot be clearly tagged as RGB or AGB. We note that these stars do not change the trends discussed in the plots, therefore we keep them in our analysis.

We note that all other stars are located at the upper RGB, that is, during the RGB bump or above it. \cite{gratton+00} discussed how the mixing that takes place at the RGB bump changes the abundances of C and N with respect to the lower RGB abundances (see also the review by \citealp{salaris+02}). These changes do not imply changes in the CN-CH distributions, as we show in this paper.

The estimation of the RGB bump magnitude was done at a first approximation based on the equation as a function of metallicity by \cite{salaris+02} that results in 0.47~mag above the horizontal branch (V $= 14.8$ mag, \citealp[][2010 edition]{harris96}). However, the empirical and theoretical values differ by about 0.2 to 0.4~mag at the 
metallicity of NGC~3201 \citep[${\rm [Fe/H]}$ = -1.46, see e.g.][]{riello+03,dicecco+10}. Therefore, it is fair to say that the RGB bump of NGC~3201 is at V $\sim14.7$~mag. This is the value estimated by \cite{kravtsov+09} with the same photometric data. In panels c and d of Fig. \ref{cmd-mask1} we show the luminosity function of all stars and RGB stars respectively, indicating the over densities of the horizontal branch and RGB bump, which agree with the expected values.

\subsection{Spectroscopic observations}
\label{sec:specobs}

After selecting the targets and preparing the masks, we performed the spectroscopic observations in Multi Object Spectroscopy (MOS) mode of EFOSC2. For each pointing (centre, north, south) we took three observations of 20 minutes each to reach enough signal-to-noise (S/N) and to correct cosmic rays.
We used grism \#07, which covers the blue region from 3270-5240$\rm \AA$, slit width 1.34$\arcsec,$ and length 8.6$\arcsec$, which means a spectral resolution of $\Delta\lambda=$ 7.4 $\rm \AA$, or R$\approx$500. The pixel scale is 0.12$\arcsec$, and we chose binned readout mode 2x2 that makes the pixel scale be 0.24$\arcsec$. Data reduction was done also using the ESO pipeline {\it esorex}. Each spectra was normalised locally to measure each index of CN and CH separately. For the CN index, we fitted a straight line to the two pseudo-continua defined by \cite{pickles85}, and for CH we proceeded in the same way using the two pseudo-continua defined by \cite{harbeck+03}.
Table \ref{log} gives further information.

\begin{table*}
\footnotesize
\begin{center}
\caption{Log of observations for pre-images and spectroscopy.}
\label{log}
\renewcommand{\arraystretch}{1.1}
\begin{tabular}{llllllllll}

\hline \hline
\noalign{\smallskip}
Obs.Type & Name & RA & DEC & Obs. date &  Obs. time &Filter/Grism & Exp. time & Airmass & Seeing ($\arcsec$) \\
\noalign{\smallskip}
\hline    
\noalign{\smallskip}
  IMA  &  NGC3201 (centre)  & 10:17:37 & -46:24:34 & 21-01-2011 &06:46:20  & B & 20 & 1.05 & 0.78 \\ 
(archive) &                   &          &           &            &06:41:10  & V & 10 & 1.05 & 0.87 \\
\noalign{\smallskip}
\hline    
\noalign{\smallskip}
  IMA  &  NGC3201 (north)  & 10:17:36 & -46:22:40 & 28-02-2016 &05:15:25  & B & 20 & 1.06 & 0.61   \\ 
       &                   &          &           &            &05:17:22  & V & 10 & 1.01 & 0.61   \\ 
       &   NGC3201 (south) & 10:17:36 & -46:26:40 & 29-02-2016 &05:21:17  & B & 20 & 1.06 &  0.58  \\
       &                   &          &           &            &05:23:14  & V & 10 & 1.06 &  0.61  \\
\noalign{\smallskip}
\hline
\noalign{\smallskip}
  MOS  &  NGC3201 (centre) & 10:17:36 & -46:24:40 & 29-02-2016 & 03:33:00  & $Gr\#7$ & 1200$\times$3 & 1.05 &  0.91 \\
  \noalign{\smallskip}
       &  NGC3201 (north)  & 10:17:36 & -46:22:40 & 29-02-2016 & 05:01:02  & $Gr\#7$ & 1800$\times$1   & 1.07 & 0.89 \\
       &  NGC3201 (north)  & 10:17:36 & -46:22:40 & 01-03-2016 & 03:38:35  & $Gr\#7$ & 1800$\times$1   & 1.05 & 0.69 \\
       \noalign{\smallskip}
       &  NGC3201 (south)  & 10:17:36 & -46:26:40 & 01-03-2016  & 02:17:25  & $Gr\#7$ & 1200$\times$3 & 1.07 & 0.69  \\
  \noalign{\smallskip}
   \hline 
      \end{tabular}
      \end{center}
      \end{table*}

%

\section{Spectroscopic analysis}

\subsection{Membership selection}

\begin{figure}
\begin{center}
\includegraphics[width = \columnwidth,angle=-90]{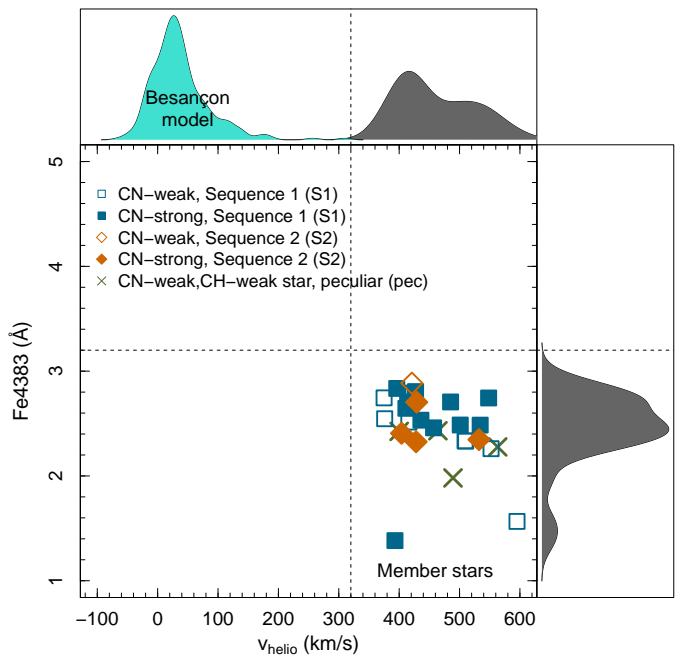}
\caption{Heliocentric velocities versus Fe4383 index for all 28 valid spectra. Smoothed histograms 
show the distribution of the two parameters. We also show the velocities distribution from the
Besan\c con model. Member stars are located in the lower right quadrant indicated by the dashed lines. 
Sequence 1 ({\it S1}), sequence 2 ({\it S2}), and peculiar stars ({\it pec.}) defined in Fig. \ref{NaO-CNCH} are indicated by blue squares, orange diamonds, and green crosses, respectively. CN-strong ($\delta$S(3839) $\geq$ 0) and CN-weak stars ($\delta$S(3839) $<$ 0) are represented by filled and open symbols, respectively. These symbols and colours are used also in the following figures. Dashed lines are set by eye at the lower limit of v$_{\rm helio}$ and the upper limit of Fe4383 distributions.}
\label{Vel_Helio_versus_Iron}
\end{center}
\end{figure}

From 48 spectra observed, we excluded six that did not have enough coverage to measure a CN index. These include the AGB star with V=13.2~mag and one of the RGB tip stars. Two stars were observed twice and we excluded the two spectra with the lower S/N. From the 40 remaining spectra, 12 were excluded because they had S/N$_{\rm CN}\leq 25$ (i.e. S/N$_{\rm CH} \leq 55$), which implies $\sigma_{CN} \geq 0.06$ and $\sigma_{CH} \geq 0.03$ as shown in Fig. \ref{snsig}. Among them is the outlier at (V,B-V) = (15.1, 1.3). We ended up with a sample of 28 valid spectra. The other three RGB tip stars are not biased to any particular group in Fig. \ref{Vel_Helio_versus_Iron}, two are {\it S1}, and the other is {\it pec}.

To ensure that we are studying stars only from NGC~3201, member stars were selected using the traditional plot of Fe abundance versus heliocentric velocity (Fig. \ref{Vel_Helio_versus_Iron}).
We derived radial velocities of the individual stars using the cross-correlation python package {\it crosscorrRV.}\footnote{\url{http://pyastronomy.readthedocs.io/en/latest/}.} This package calculates the radial velocity of stars shifting the rest-frame wavelength axis of a template. We used as template a synthetic spectrum (\citealp{coelho+05}\footnote{\url{http://specmodels.iag.usp.br/}.}) for a typical red giant star in NGC 3201 with parameters T$_{\rm eff} = 5000$ K, log($g$)=1.0, [Fe/H]$=-1.5$, and [$\alpha$/Fe]$=+0.4$. The resolution of the synthetic spectrum ($\Delta\lambda$ $\approx$ 0.2) was degraded using IRAF task {\it gauss} to match the resolution of the stellar spectra of our sample ($\Delta\lambda$ $\approx$ 7.4). We performed cross correlation using the K and H Calcium lines region (3933.66 \AA \ and 3968.47 \AA). We also corrected the radial velocities from the motion of the Earth to the heliocentric reference using the python package {\it helcorr}$^2$. Errors were assumed to be the full width at half maximum (FWHM) of the cross-correlation function, which is about 5~km/s for all spectra.
Fe abundance is indicated by the Lick index Fe4383 measured with the LECTOR code.\footnote{\url{http://www.iac.es/galeria/vazdekis/vazdekis_software.html}} 

We ran a Bensan\c con model\footnote{\url{model.obs-besancon.fr}} \citep{robin+03} with a solid angle of 0.2 deg$^2$ at the direction of NGC~3201 and selecting only giant, bright giant, and supergiant stars from all Galactic components with visual magnitude between 18 $<$ V $<$ 10. The simulation resulted in 309 stars with metallicities higher than [Fe/H] $\gtrsim$ -1.5, which is the metallicity of NGC~3201 \citep{dias+16msgr}\footnote{\url{www.sc.eso.org/~bdias/catalogues.html}}; only three stars had a metallicity below that. Radial velocities are indicated in Fig. \ref{Vel_Helio_versus_Iron} and show 
no overlap.
Therefore, field stars would have v$_{\rm helio} < $320 km/s and higher metallicities, that is they would be in the upper left quadrant, where there are no stars.
In other words, all 28 valid stars were selected as members. The average velocity of the member stars is v$_{\rm helio} =$ 460$\pm$63 km/s, which is compatible with 494km/s from \citet[][2010 edition]{harris96}.

\subsection{Definition of CN and CH indices}

We adopt here the modified index definition of CN (S3839) and CH (G4300) by \cite{harbeck+03} to have a homogeneous analysis and compare results with those from \cite{kayser+08} who used Harbeck's definition. 
The indices defined by \cite{harbeck+03} have slightly different spectral regions with respect to the classical ones defined by \cite{norris+81} in order to avoid strong hydrogen lines nearby in main sequence stellar spectra. This was not an issue for the RGB stars observed in the 1980s. The differences in the indices using different definitions are briefly discussed in Sect. \ref{sec:cndist}.

\begin{figure}[!htb]
\includegraphics[angle=-90,width=\columnwidth]{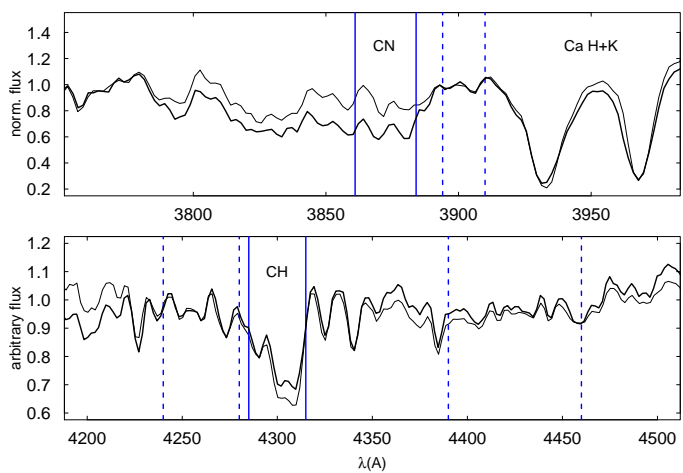}
\caption{Definition of CN and CH indices S(3839) and G4300 from Eqs. \ref{eqCN} and \ref{eqCH} using spectra of two NGC~3201 stars. Thin lines are from the CN-weak, CH-strong star N3201centre\_08, thick lines are from the CN-strong, CH-weak star N3201centre\_11, both {\it S1} stars with V$\approx$13 mag. The index region is highlighted by solid blue lines and pseudo-continuum regions are identified by dashed blue lines. Spectra were locally normalised as explained in Sect. \ref{sec:specobs}.
}
\label{figdef}
\end{figure}

Indices from Equations \ref{eqCN} and \ref{eqCH} were measured using an R code\footnote{\url{https://www.r-project.org/} \citep{R}} written by B. Dias (see Fig. \ref{figdef}) resulting in

\begin{equation}
{\rm S(3839)} = -2.5 \cdot {\rm log}\left(\dfrac{\int_{3861}^{3884}F_{\lambda}d\lambda}{\int_{3894}^{3910}F_{\lambda}d\lambda}\right)
\label{eqCN}
\end{equation}

\begin{equation}
{\rm G4300} = -2.5 \cdot {\rm log}\left(\dfrac{\int_{4285}^{4315}F_{\lambda}d\lambda}{0.5 \cdot \int_{4240}^{4280}F_{\lambda}d\lambda + 0.5 \cdot  \int_{4390}^{4460}F_{\lambda}d\lambda } \right).
\label{eqCH}
\end{equation}

\noindent Uncertainties are discussed in Appendix \ref{appError}.

\subsection{Surface temperature and gravity effects}
\label{sec:tefflogg}

\begin{figure}[!htb]
\centering
\includegraphics[angle=-90,width=0.78\columnwidth]{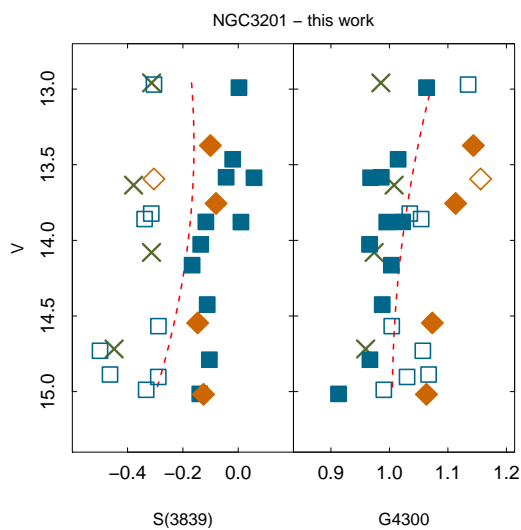}
\caption{Correction of S(3839) and G4300 for stellar surface temperature and gravity for 28 member and good quality (S/N$_{\rm CN} \geq 25$) RGB stars of NGC~3201.  We fitted a second-order polynomial to each dataset, which is represented by the red dashed lines. The difference between the indices and the fitted line for a given magnitude defines $\delta$S(3839) and $\delta$G4300. Symbols are the same as in Fig.\ref{Vel_Helio_versus_Iron}.
If the peculiar stars indicated by the crosses are not considered, two groups of stars with anti-correlated CN and CH can be identified: {\it S1}, indicated by squares, and {\it S2}, indicated by diamonds 
(see also Fig. \ref{NaO-CNCH} and discussions). Error bars are of the order of the point size, and are thus omitted in the plots.}
\label{fig:tefflogg}
\end{figure}

Both S(3839) and G4300 indices are sensitive to surface gravity and effective temperature \citep[see][]{norris+81}. 
Removing this dependency is a {\it sine qua non} condition for detecting their sensitivity to nitrogen and carbon.
The typical proxies used to correct the indices are colour or magnitude \citep{norris+81,harbeck+03,kayser+08,campbell+12}. We adopt here a similar method to \cite{pancino+10} who traced a ridge line in the distribution of index versus magnitude. Instead of a ridge line we fitted a second-order polynomial to this distribution;
Figure \ref{fig:tefflogg} displays the polynomial fit to the data. The difference between the index and the polynomial 
for a given magnitude defines the excess indices $\delta$S(3839) and $\delta$G4300 that will be used in this paper from now on. We note that the spread seen in Fig. \ref{fig:tefflogg} is real because all member stars have $\sigma_{CN} < 0.06$ and $\sigma_{CH} < 0.03$. The parameters for all member stars are given in Table \ref{finalparam}.

\begin{table*}
\scriptsize
\begin{center}
\caption{Final parameters for all 28 good quality member stars of NGC~3201}
\label{finalparam}
\renewcommand{\arraystretch}{1.1}
\scriptsize
\begin{tabular}{llllllllrrrrr}

\hline \hline
\noalign{\smallskip}
Star & RA & DEC & V &  B-V  &   v$_{\rm helio}$   &   Fe4383   &  S/N$_{\rm CN}$  &  S/N$_{\rm CH}$  &   CN=S(3839)  &   CH=G4300 &    $\delta$CN   &   $\delta$CH \\
 & hh:mm:ss.sss & dd:mm:ss.sss & mag &  mag  &  km/s   &   {\rm \AA} &   &     &   mag &   mag &    mag   &   mag \\
(1)  & (2) & (3) & (4) & (5)  &  (6)  & (7)   &   (8) & (9)  & (10)   & (11)    & (12) & (13) \\
\noalign{\smallskip}
\hline    
\noalign{\smallskip}

  N3201centre\_01  &  10:17:49.057 & -46:24:58.572  &  13.823 & 1.186  &  375 & 2.745  & 60.37 & 129.7  &  -0.314 $\pm$ 0.026  &  1.035 $\pm$ 0.012  &  -0.144  &   0.006  \\
  N3201centre\_02  &  10:17:47.474 & -46:23:53.786  &  12.959 & 1.287  &  400 & 2.425  & 79.79 & 165.5  &  -0.313 $\pm$ 0.020  &  0.985 $\pm$ 0.009  &  -0.144  &  -0.088  \\
  N3201centre\_03  &  10:17:45.289 & -46:24:27.292  &  13.878 & 1.142  &  436 & 2.532  & 53.99 & 116.9  &  -0.117 $\pm$ 0.031  &  1.023 $\pm$ 0.013  &   0.056  &  -0.004  \\
  N3201centre\_04  &  10:17:44.273 & -46:23:43.753  &  14.903 & 1.039  &  416 & 2.514  & 31.81 & 64.89  &  -0.289 $\pm$ 0.050  &  1.030 $\pm$ 0.023  &  -0.007  &   0.025  \\
  N3201centre\_05  &  10:17:43.300 & -46:24:38.876  &  13.756 & 1.160  &  428 & 2.326  & 51.93 & 117.5  &  -0.080 $\pm$ 0.032  &  1.113 $\pm$ 0.013  &   0.088  &   0.082  \\
  N3201centre\_06  &  10:17:41.993 & -46:23:54.265  &  15.015 & 0.999  &  393 & 1.383  & 49.58 & 80.20  &  -0.139 $\pm$ 0.033  &  0.913 $\pm$ 0.018  &   0.161  &  -0.092  \\
  N3201centre\_07  &  10:17:40.756 & -46:24:44.838  &  13.594 & 1.207  &  421 & 2.885  & 56.02 & 119.2  &  -0.305 $\pm$ 0.028  &  1.156 $\pm$ 0.013  &  -0.142  &   0.118  \\
  N3201centre\_08  &  10:17:38.906 & -46:24:02.333  &  12.970 & 1.259  &  423 & 2.791  & 52.22 & 117.3  &  -0.305 $\pm$ 0.030  &  1.135 $\pm$ 0.013  &  -0.136  &   0.062  \\
  N3201centre\_09  &  10:17:37.480 & -46:25:02.046  &  13.373 & 1.224  &  429 & 2.704  & 50.29 & 121.3  &  -0.101 $\pm$ 0.033  &  1.143 $\pm$ 0.013  &   0.060  &   0.095  \\
  N3201centre\_10  &  10:17:36.040 & -46:24:02.707  &  14.027 & 1.089  &  396 & 2.835  & 28.54 & 62.61  &  -0.136 $\pm$ 0.058  &  0.966 $\pm$ 0.023  &   0.047  &  -0.055  \\
  N3201centre\_11  &  10:17:34.118 & -46:24:14.861  &  12.990 & 1.273  &  414 & 2.707  & 33.71 & 89.90  &   0.003 $\pm$ 0.051  &  1.063 $\pm$ 0.017  &   0.170  &  -0.008  \\
  N3201centre\_15  &  10:17:29.949 & -46:25:30.565  &  13.859 & 1.134  &  376 & 2.546  & 39.99 & 111.3  &  -0.338 $\pm$ 0.039  &  1.054 $\pm$ 0.014  &  -0.166  &   0.027  \\
  N3201north\_01   &  10:17:45.818 & -46:23:27.370  &  14.546 & 1.085  &  404 & 2.407  & 27.31 & 62.42  &  -0.147 $\pm$ 0.060  &  1.073 $\pm$ 0.024  &   0.085  &   0.064  \\
  N3201north\_03   &  10:17:42.926 & -46:22:48.306  &  13.586 & 1.188  &  457 & 2.459  & 35.43 & 80.76  &   0.057 $\pm$ 0.048  &  0.968 $\pm$ 0.019  &   0.219  &  -0.071  \\
  N3201south\_02   &  10:17:46.611 & -46:26:15.954  &  13.879 & 1.162  &  427 & 2.805  & 71.33 & 159.7  &   0.010 $\pm$ 0.024  &  0.995 $\pm$ 0.009  &   0.183  &  -0.031  \\
  N3201south\_03   &  10:17:45.398 & -46:25:31.757  &  13.635 & 1.214  &  464 & 2.431  & 106.5 & 191.6  &  -0.378 $\pm$ 0.015  &  1.008 $\pm$ 0.008  &  -0.214  &  -0.028  \\
  N3201south\_04   &  10:17:44.231 & -46:27:16.916  &  14.789 & 1.074  &  501 & 2.486  & 72.59 & 142.0  &  -0.104 $\pm$ 0.023  &  0.967 $\pm$ 0.011  &   0.160  &  -0.039  \\
  N3201south\_06   &  10:17:41.264 & -46:26:48.642  &  14.988 & 1.048  &  509 & 2.334  & 64.56 & 104.7  &  -0.333 $\pm$ 0.024  &  0.990 $\pm$ 0.014  &  -0.037  &  -0.015  \\
  N3201south\_07   &  10:17:40.303 & -46:26:13.456  &  14.424 & 1.027  &  411 & 2.645  & 54.16 & 100.9  &  -0.112 $\pm$ 0.030  &  0.988 $\pm$ 0.015  &   0.105  &  -0.023  \\
  N3201south\_08   &  10:17:39.393 & -46:26:58.362  &  14.730 & 1.076  &  552 & 2.260  & 53.58 & 122.4  &  -0.500 $\pm$ 0.028  &  1.057 $\pm$ 0.012  &  -0.244  &   0.050  \\
  N3201south\_09   &  10:17:37.655 & -46:26:12.836  &  13.464 & 1.231  &  485 & 2.706  & 84.15 & 180.1  &  -0.020 $\pm$ 0.020  &  1.015 $\pm$ 0.008  &   0.141  &  -0.029  \\
  N3201south\_10   &  10:17:36.541 & -46:27:38.682  &  14.567 & 1.007  &  510 & 2.333  & 64.45 & 106.6  &  -0.288 $\pm$ 0.024  &  1.004 $\pm$ 0.014  &  -0.054  &  -0.005  \\
  N3201south\_11   &  10:17:34.490 & -46:25:36.160  &  14.718 & 1.027  &  489 & 1.981  & 75.59 & 125.5  &  -0.449 $\pm$ 0.020  &  0.959 $\pm$ 0.012  &  -0.194  &  -0.047  \\
  N3201south\_12   &  10:17:33.376 & -46:27:36.115  &  14.887 & 0.974  &  595 & 1.566  & 55.55 & 166.3  &  -0.463 $\pm$ 0.027  &  1.067 $\pm$ 0.009  &  -0.184  &   0.061  \\
  N3201south\_13   &  10:17:32.096 & -46:25:09.937  &  14.164 & 1.073  &  548 & 2.744  & 70.67 & 125.7  &  -0.167 $\pm$ 0.022  &  1.003 $\pm$ 0.012  &   0.026  &  -0.014  \\
  N3201south\_14   &  10:17:30.029 & -46:26:36.726  &  14.080 & 1.093  &  563 & 2.277  & 83.27 & 185.7  &  -0.313 $\pm$ 0.018  &  0.974 $\pm$ 0.008  &  -0.127  &  -0.046  \\
  N3201south\_16   &  10:17:26.975 & -46:25:52.468  &  13.583 & 1.175  &  534 & 2.485  & 72.38 & 145.2  &  -0.044 $\pm$ 0.023  &  0.986 $\pm$ 0.010  &   0.119  &  -0.053  \\
  N3201south\_18   &  10:17:26.013 & -46:25:18.181  &  15.018 & 1.015  &  532 & 2.347  & 60.06 & 106.7  &  -0.126 $\pm$ 0.028  &  1.063 $\pm$ 0.014  &   0.175  &   0.058  \\

   \hline 
      \end{tabular}
      \end{center}
      \tablefoot{ 
\tablefoottext{1}{Star ID from our three masks.}
\tablefoottext{2,3}{Coordinates of each star, equinox J2000.0.}
\tablefoottext{4,5}{Magnitude and colour from \cite{kravtsov+09} calibrated and corrected by differential reddening.}
\tablefoottext{6}{Heliocentric velocities. Errors are about 5~km/s, assumed as the FWHM of the cross-correlation function.}
\tablefoottext{7}{Lick index of iron used as proxy for metallicity.}
\tablefoottext{8,9}{Signal-to-noise ratio at the wavelength of CN and CH indices.}
\tablefoottext{10,11}{CN and CH indices as defined in Equations \ref{eqCN} and \ref{eqCH}.}
\tablefoottext{12,13}{CN and CH indices corrected by surface gravity effects as discussed in Sect. \ref{sec:tefflogg}.}
}
      \end{table*}

%
\section{Multiple stellar generations in NGC~3201}
\label{sec:multipop}

\subsection{The unusual CN-CH anti-correlation of NGC~3201}
\label{sec:highresspec}

\begin{figure}[!htb]
\centering
\includegraphics[angle=-90,width=\columnwidth]{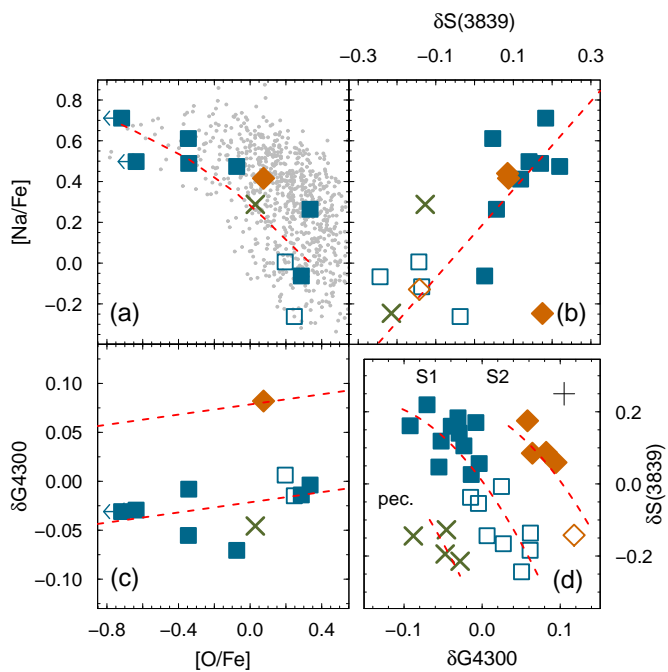}
\caption{ {\bf (a)} Na-O anti-correlation for all globular clusters from \cite{carretta+09gir} in grey dots superimposed by the 11 NGC~3201 stars in common with our sample with available [Na/Fe] and [O/Fe]; symbols are the same as in Fig. \ref{Vel_Helio_versus_Iron}. Smooth spline function is shown
as red dashed line to highlight the anti-correlation in our data. Peculiar stars indicated by green crosses 
are not considered in the fitted function based on the selection of panel (d). 
{\bf (b)} Correlation between sodium and cyanogen, which is a proxy for nitrogen abundances for the 18 stars in common with available [Na/Fe]. A straight line is fitted to {\it S1} stars.
{\bf (c)} Same as (b) but for CH and oxygen for stars in common with available [O/Fe]. The fitted line is shifted by 0.1~mag in $\delta$G4300 to match the only {\it S2} star.
{\bf (d)}  Distribution of CN and CH indices for all 28 NGC~3201 good quality member stars, revealing three groups of stars: sequence 1 ({\it S1}), sequence 2 ({\it S2}), and peculiar ({\it pec}).
A second-order polynomial was fitted to \textit{\textbf{S1}} stars and the same curve was shifted by 0.1~mag in $\delta$G4300 as in panel (c) to match the \textit{\textbf{S2}} stars, and by $-$0.1~mag to match the peculiar stars. 
Mean uncertainties are shown by the cross on the upper right corner of the panel.}
\label{NaO-CNCH}
\end{figure}

We show in Fig. \ref{NaO-CNCH}d the anti-correlation between the corrected indices $\delta$S(3839) and $\delta$G4300 for the 28 good quality member stars of NGC~3201. We split them into two CN-CH anti-correlation sequences, namely {\it S1} and {\it S2,} and a group of CN-weak, CH-weak peculiar ({\it pec}) stars. We fitted a second-order polynomial to {\it S1} and shifted the curve to match {\it S2} and peculiars. The separation between the sequences is $\Delta\delta$G4300 $=$ 0.1~mag, which is about 7$\sigma$ distance. Therefore the separation of our sample into these groups is significative and we adopt the same point colours and shapes defined in Fig. \ref{NaO-CNCH}d to the other plots in this paper.
One peculiar star was already indicated by \cite{smith+82} and \cite{dacosta+81} and we now increase this sample to four peculiar stars. The other panels of Fig. \ref{NaO-CNCH} show that stars from different groups are indistinguishable in the Na-O anti-correlation, at least with the limited sample we have. The usual correlations Na-CN and O-CH are also shown, where first (1G, Na-poor, CN-weak) and second generation (2G, Na-rich, CN-strong) stars can be identified.

The globular cluster M~22 also has two sequences of C-N anti-correlation and the C-rich sequence is less populous 
than the C-poor sequence as in the case of {\it S2} in comparison with {\it S1} for NGC~3201. In the case of M~22, s-process element abundances were available and the conclusion of \cite{marino+11} was that the C-poor/N-poor sequence (equivalent to {\it S1}) is s-poor and Fe-poor, while the C-rich/N-rich sequence (equivalent to {\it S2}) is s-rich and Fe-rich. Oxygen is the same for the two groups, but C+N+O is higher for the s-rich stars. High-resolution spectroscopy is needed to confirm whether NGC~3201 has an increasing C+N+O and s-process element abundances in the sequence {\it S1-S2} as well.
\cite{gonzalez+98} found hints that O-rich NGC~3201 stars are depleted in [Ba/Eu]. In Fig. \ref{NaO-CNCH} we show that O-rich stars are C-rich, which would mean that the sequence {\it pec}-{\it S1}-{\it S2} has decreasing s-element abundances, not increasing as is the case for M~22. However, 11 out of the 13 stars studied by Gonzalez \& Wallerstein have constant [Ba/Eu] abundance ratios, and only two O-rich stars are depleted in Ba. A larger sample, preferably including the {\it pec}-{\it S1}-{\it S2} stars, is needed in order to derive abundances of C, N, O, and s-/r-process elements.

 \cite{lim+17} analysed the double CN-CH anti-correlation of M~22 in a different way. They argued that these stars show a positive CN-CH correlation, which is also the case for other anomalous clusters
 such as NGC~1851, NGC~6273, and NGC~5286. They showed that the CN-strong stars are s-rich, and CN-weak stars are s-poor. NGC~288 was used as reference and it shows only a single anti-correlation. If the CN-CH relation of NGC~3201 stars is interpreted following this line, it is another indication that the sequence {\it pec}-{\it S1}-{\it S2} may have increasing s-element abundances and possibly increasing metallicities.

The CN-CH anti-correlation can be explained by CN-cycle processing. Some clusters show that C+N increases with decreasing C \citep[e.g. M~5;][]{cohen+02} and ON-cycle processed material is required to explain that.
During the first dredge-up, the convective envelope of an RGB star can move material processed during the CNO cycle up to the atmosphere. However, for metal-poor stars, the convective envelope does not go deep enough to reach the H-burning shell to mix enhanced (or depleted) CNO-cycle processed elements.\footnote{For a recent review, see \cite{gratton+04} and references therein.} NGC~3201 is relatively metal-poor, but {\it S1}, {\it S2}, and peculiar stars have different CH strength (proxy for C) and we can only speculate that these three groups may have different C+N abundances.

\subsection{Seeking a photometric index that splits {\it pec}-{\it S1}-{\it S2} stars}

\cite{monelli+13} defined a colour index based on Johnson filters U, B, I as given 
by the equation

\begin{equation}
{\rm c}_{\rm U,B,I} = {\rm (U-B) - (B-I).}
\label{cubi}
\end{equation}

\noindent They found that the V-c$_{\rm U,B,I}$ diagram reveals a split RGB for all globular clusters analysed, after differential reddening correction. The multiple branches correlate with the chemical abundances of light-elements O, Na, C, N, and Al.

\begin{figure}[!htb]
\centering
\includegraphics[angle=-90,width=0.86\columnwidth]{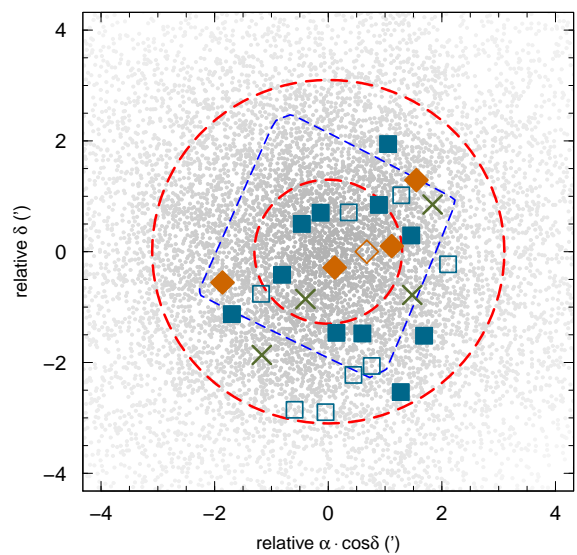}
\caption{Spatial distribution of NGC~3201 stars from \cite{kravtsov+09} in grey dots with a greyscale density pattern. Spectroscopic targets are shown with the same symbols as in Fig. \ref{Vel_Helio_versus_Iron} for the 28 good quality member stars.
Dashed outer circle shows the half-light radius (r$_{\rm h}=3.1\arcmin$),
the internal dashed red circle is the core radius (r$_{\rm c}=1.3\arcmin$), and the blue dashed square delimits the region of the HST observations used here.}
\label{spatialdist}
\end{figure}

\begin{figure}[!htb]
\centering
\includegraphics[width=0.7\columnwidth]{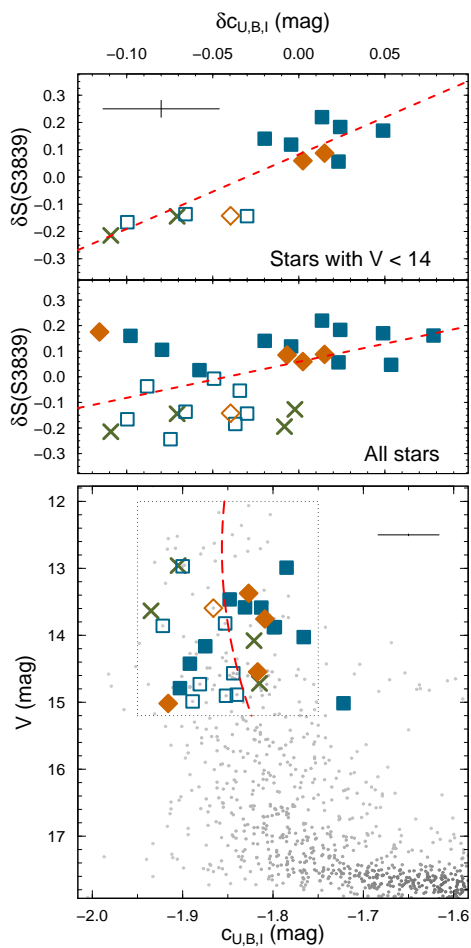}
\caption{{\bf Bottom panel:} V-c$_{\rm U,B,I}$ diagram for NGC~3201 using photometry from \cite{kravtsov+09} corrected by differential reddening but without decontamination from field stars. 
We selected a rectangle around the RGB stars and fitted a second-order polynomial shown by a dashed red line.
This line is used to define $\delta$c$_{\rm U,B,I}$ as the difference between c$_{\rm U,B,I}$ and the fitted line for a given V magnitude for the 28 stars analysed here. Symbols are the same as in Fig. \ref{Vel_Helio_versus_Iron}. Average error bars are indicated at the upper right corner.
{\bf Middle panel:} Correlation between $\delta$S(3839) and $\delta$c$_{\rm U,B,I}$ for all 28 stars. A linear fit to the {\it S1} stars is shown by a red dashed line.
{\bf Upper panel:} Same as middle panel but only for the 14 stars brighter than V $<$ 14~mag. Average error bars are indicated at the upper left corner.}
\label{V-CUBI}
\end{figure}

We adopted the available UBVI photometry corrected by differential reddening by \cite{kravtsov+09} to
produce a V-c$_{\rm U,B,I}$ diagram for NGC~3201. In Fig. \ref{spatialdist} we show the sky distribution of the stars from \cite{kravtsov+09} and identify the core radius and half-light radius, as well as our 28 targets.
We plotted the pseudo-CMD, aka V-c$_{\rm U,B,I}$ diagram, using stars from 
Fig. \ref{spatialdist} and displayed them on Fig. \ref{V-CUBI}.
We also fitted a 
second-order polynomial to all RGB stars within a box defined by $12<$V$<16$~mag and $-2.0<$c$_{\rm U,B,I}<-1.7$~mag.
In a similar way to what it was done to correct S(3839) to $\delta$S(3839), we define $\delta$c$_{\rm U,B,I}$ as the difference between c$_{\rm U,B,I}$ and the fitted line for a given V magnitude.

The corrected spectroscopic index $\delta$S(3839)  is plotted against the corrected photometric index $\delta$c$_{\rm U,B,I}$ on the upper panels of Fig. \ref{V-CUBI}.
Stars fainter than V$>$14~mag have larger photometric uncertainties and generate dispersion on the plot. If only bright stars are plotted, a correlation between these indices is clearly seen, as expected, and 1G and 2G stars are well separated. However, {\it pec}-{\it S1}-{\it S2} stars are all mixed. In conclusion, $\delta$c$_{\rm U,B,I}$ cannot split the three groups.

UV filters are also useful to split the RGB of globular clusters.
The Hubble Space Telescope (HST) Large Legacy Treasury Program \citep{piotto+15} had its first public data release with the homogeneous photometric catalogues published recently \citep{soto+17}. This release is preliminary, therefore no differential reddening correction was taken into account, which could cause some
dispersion on the photometric indices but not as much as ground-based photometry.
This team defined the colour index 

\begin{equation}
{\rm c}_{\rm F275W,F336W,F438W} = (m_{\rm F275W} - m_{\rm F336W}) - (m_{\rm F336W} - m_{\rm F438W})
\label{cuv}
\end{equation}

\noindent that is very sensitive to nitrogen abundances and is  
useful to disentangle 1G and 2G RGB stars in globular clusters.

The lower left panel of Fig. \ref{V-CUBI} shows the V-$\delta$c$_{\rm F275W,F336W,F438W}$ diagram where
the {\it S2} CN-strong stars fall in between the CN-strong and CN-weak stars of {\it S1}, which is a possible explanation for why the RGB is not clearly bimodal 
but rather has a smooth transition. 
If {\it S2} stars are indeed s-rich and {\it S1} s-poor, the RGB in this CMD would have s-poor stars in the left branch 
and mixed s-poor and s-rich stars towards the right branch. We notice that the {\it S1}-1G star to the right is likely to be there because it has $\delta$S(3839) very close to zero and is therefore more sensitive to our definition of CN-strong and CN-weak with a cut at $\delta$S(3839)~$ = 0$. This CMD seems useful to split {\it S1}-2G from the others.
As done for c$_{\rm U,B,I}$ we also fitted a line to the RGB stars and defined the differential index $\delta$c$_{\rm F275W,F336W,F438W}$ as the difference between c$_{\rm F275W,F336W,F438W}$ and the fitted line for a given magnitude. The correlation between $\delta$c$_{\rm F275W,F336W,F438W}$ and $\delta$S(3839) on the upper left panel is clear, as expected. It splits 1G and 2G stars but mixes {\it S1} and {\it S2} stars. Interestingly, peculiar stars seem to have lower $\delta$c$_{\rm F275W,F336W,F438W}$ values for the same $\delta$S(3839) as the 1G stars. This correlation seems useful to find peculiar stars.

\begin{figure}[!htb]
\centering
\includegraphics[angle=-90,width=\columnwidth]{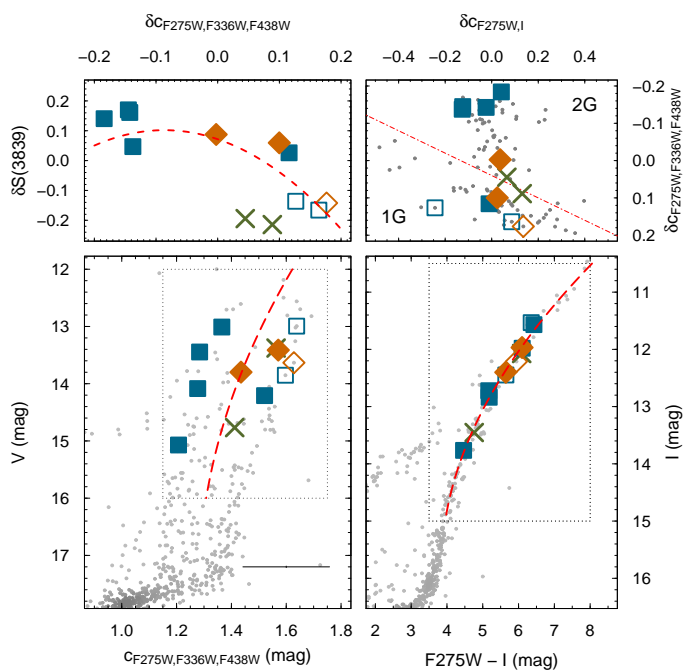}
\caption{Similar to Fig. \ref{V-CUBI} but using HST UV photometry from \cite{soto+17}, available only for 12 out of the 19 good quality member stars within the HST field of view (See Fig. \ref{spatialdist}). The bottom panels show the CMDs with our spectroscopic targets identified using the same symbols as in Fig. \ref{Vel_Helio_versus_Iron} . 
The index $\delta$c$_{\rm F275W,F336W,F438W}$ is plotted against $\delta$S(3839) on the upper left panel.
A second-order polynomial was fitted to {\it S1} stars, and it is shown as red dashed line.
The index $\delta$c$_{\rm F275W,F336W,F438W}$ is plotted also against $\delta$c$_{\rm F275W,I}$ on the upper right panel, which is a version of the so-called `chromosome map' from \cite{milone+17} that distinguishes 1G and 2G stars, split by a line inclined by $\theta = 18^{\circ}$. Our spectroscopic targets are identified on the map.
}
\label{hst-cmd}
\end{figure}

\cite{milone+17} defined the so-called `chromosome map' that consists of a plot of the pseudo colour index $\delta$c$_{\rm F275W,F336W,F438W}$ versus the colour index $\delta$c$_{\rm F275W,F814W}$. On the first data release of the HST UV treasury, the F814W magnitude was not yet available, but the authors provide an I magnitude which is very similar. We replaced $m_{\rm F814W}$ by I and produced our version of the `chromosome map' as shown on the upper right panel of Fig. \ref{hst-cmd}. The regions of 1G and 2G stars are clearly identified and split by a line inclined by $\theta=18^{\circ}$ as defined by Milone et al.. Stars of type {\it S1}  follow the 1G-2G split (with the outlier being explained above). Stars of type {\it S2}  are not consistent with an additional 1G or 2G sequence shifted from the main 1G or 2G regions, as  is the case for other anomalous clusters (type II in the nomenclature of Milone et al.) such as M~22 and NGC~1851. In fact, Milone et al. classified NGC~3201 as type I. In conclusion, the chromosome map is not able to split {\it pec}-{\it S1}-{\it S2} stars.

A third photometric index was defined by \cite{marino+15} 
based on the Johnson filters B, V, I as given by 

\begin{equation}
 {\rm c}_{\rm B,V,I} = {\rm (B-V) - (V-I) .}
\label{eq:cBVI}
\end{equation}

\noindent These authors say that $ {\rm c}_{\rm B,V,I}$ 
is not very sensitive to Na or N, but it was useful to separate s-rich and s-poor stars on NGC~5286. Although s-process element abundances do not directly affect broad-band filters, at least indirectly this index was able to split well the two groups of stars. They further claim that C+N+O abundances may also affect this index. In fact,  s-rich stars are also rich in C+N+O in anomalous clusters.
We use again the photometry from \cite{kravtsov+09} to compare the photometric index with our spectroscopic measurements.

\begin{figure}[!htb]
\centering
\includegraphics[width=0.8\columnwidth]{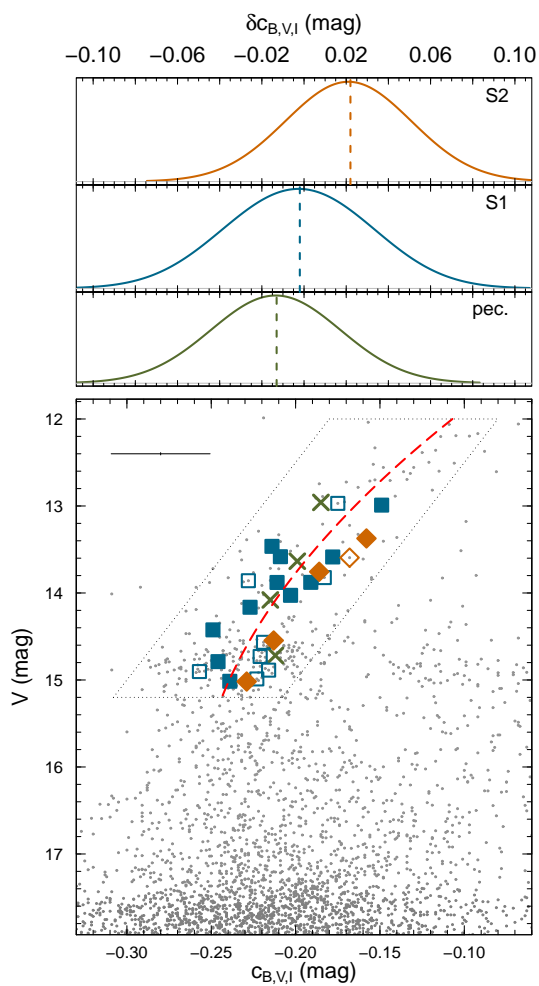}
\caption{{\bf Bottom panel:} Same as Fig. \ref{V-CUBI} but for the index c$_{\rm B,V,I}$ defined in Eq. \ref{eq:cBVI}. Average error bars are indicated at the upper left corner.  {\bf Top panels:} The KDEs of $\delta$c$_{\rm B,V,I}$ are shown for each group of stars: {\it S1}, {\it S2}, and peculiar only for stars brighter than V $<$ 14. The KDE bandwidth is the average uncertainty of 0.03. Peak values are indicated by vertical dashed lines.}
\label{V-CBVI}
\end{figure}

We show in Fig. \ref{V-CBVI} a similar CMD to the one done in Fig. \ref{V-CUBI} but now with the colour defined by Eq. \ref{eq:cBVI}. We defined the differential index $\delta$c$_{\rm B,V,I}$ in a similar way to that  for $\delta$c$_{\rm U,B,I}$.
Only stars brighter than V $<$ 14 are analysed as before.  All {\it S2} stars are to the right of the fitted line, the {\it pec} are to the left, and {\it S1} stars are spread around the curve. Kernel density estimations (KDEs) were produced for these stars separated into {\it pec}-{\it S1}-{\it S2}, and bandwidth as the average uncertainty. Although the KDEs are broad, it is possible to identify that the peaks follow the sequence {\it pec}-{\it S1}-{\it S2} in terms of increasing $\delta$c$_{\rm B,V,I}$. In conclusion, $\delta$c$_{\rm B,V,I}$ is able to split {\it pec}-{\it S1}-{\it S2} NGC~3201 stars, even if not as clearly as for NGC~5286. If this index is able to split groups of stars with different C+N+O abundance and possibly s-process element abundances, NGC~3201 could join the group of anomalous clusters.

\subsection{CN distribution}
\label{sec:cndist}

\cite{campbell+12} revealed a quadrimodal CN distribution for the anomalous cluster NGC~1851.
It is known that the more metal-rich the cluster the larger range its CN distribution has \citep[e.g.][]{kayser+08,schiavon+17,milone+17}, therefore Campbell et al. compared NGC~1851 with NGC~288 that has a similar metallicity \citep[${\rm [Fe/H] = -1.19},$][]{dias+16msgr}. They concluded that both clusters indeed have similar ranges on the CN distribution, but NGC~288, taken as a typical globular cluster, has a clearly bimodal distribution, while NGC~1851 has a quadrimodal distribution. Moreover, they showed that the two CN-strong peaks are also Ba-rich, and the other two peaks are Ba-poor. We discuss whether a quadrimodal CN distribution can be another indicator of anomalous clusters.

\begin{figure}[!htb]
\centering
\includegraphics[width=\columnwidth]{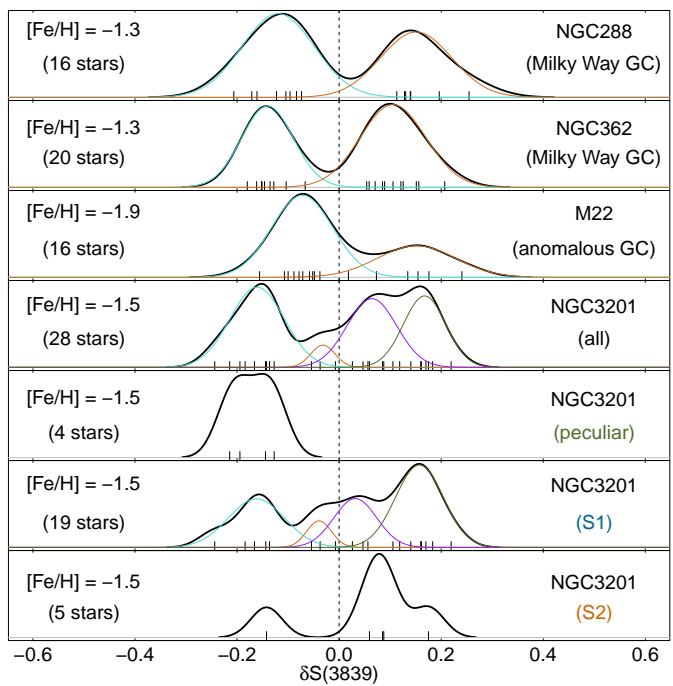}
\caption{KDE of the excess CN-index $\delta$S(3839) for NGC~3201 in comparison with two normal Galactic GCs (NGC~288, NGC~362) and one anomalous (M~22), as indicated in the panels. Data for the reference clusters were taken from \cite{kayser+08} and put in a standard scale. 
The rug plot indicates the positions of all stars in each panel. 
The KDE bandwidth is the average uncertainty of each dataset, i.e., $\sigma_{\rm NGC288}=0.055$, $\sigma_{\rm NGC362}=0.042$, $\sigma_{\rm M22}=0.046$, $\sigma_{\rm NGC3201}=0.031$.
Multi-gaussian fitting was performed to each distribution, except {\it pec} and {\it S2} stars, as discussed in the text. We also indicate cluster metallicities in the scale of \cite{dias+16}. }
\label{dcn}
\end{figure}

\begin{figure}[!htb]
\centering
\includegraphics[width=\columnwidth]{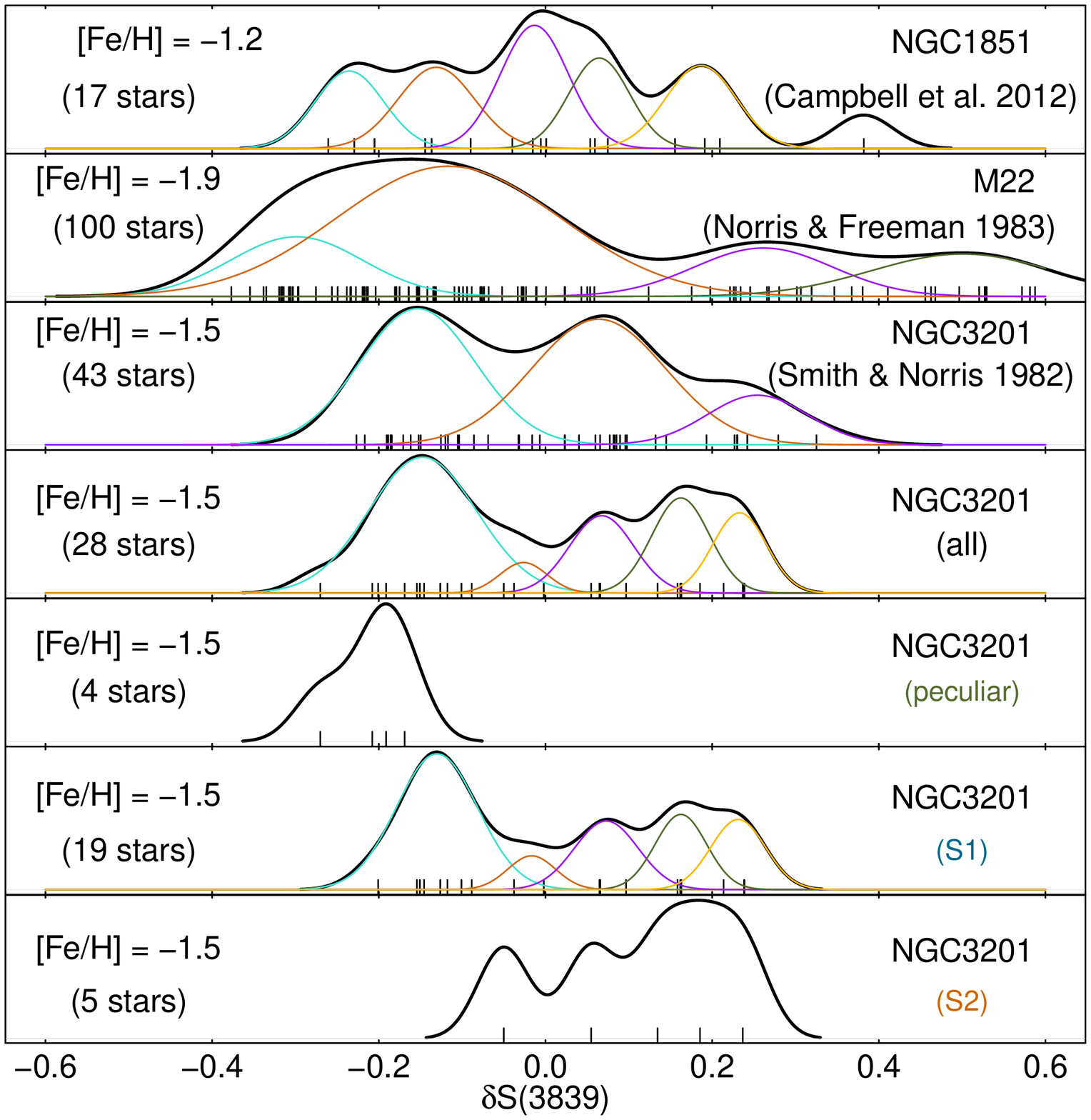}
\caption{Same as Fig. \ref{dcn} but with S(3839) measured following the classical definition used by \cite{campbell+12}, \cite{norris+83}, and \cite{smith+82} for a fair comparison with their results on NGC~1851, M~22, and NGC~3201, respectively.The  KDE bandwith is the average uncertainty of each dataset, i.e., $\sigma_{\rm NGC1851}=0.035$, $\sigma_{\rm M22}=0.07$, and $\sigma_{\rm NGC3201}=0.05$, respectively. The KDE bandwith for our dataset is 0.031.
$\delta$S(3839) was calculated for all samples following the same strategy as done in Fig. \ref{fig:tefflogg}.}
\label{cnSN82}
\end{figure}

In order to compare NGC~3201 with other clusters, we produced the KDE for each cluster shown in Fig. \ref{dcn} using the indices from the RGB stars of \cite{kayser+08} corrected following Appendix \ref{appK08}.
The KDE smoothing bandwidth was considered as the average index
uncertainty of each cluster.
For each distribution we fitted the minimum number of Gaussian functions that was needed to converge the fit using non-linear least squares.\footnote{R script adapted from \url{http://research.stowers.org/mcm/efg/R/Statistics/MixturesOfDistributions/index.htm}.} They are shown in the figure and represent well the KDEs. The sigma of each Gaussian was forced to be similar to the average index uncertainty in each case.
A possible check of the significance of the multi-gaussian fit is the number of degrees of freedom. Each Gaussian has two variables to be fit, therefore at least three points per Gaussian are needed in order to give flexibility for the fit to proceed. For NGC~288, two
Gaussians were fitted to 16 points, which is more than six, hence the result does not over-fit the KDE. The same
applies for all other clusters.

The CN distribution for all 28 good quality member RGB stars from our NGC~3201 sample shows 
four peaks (see Fig. \ref{dcn}).
This is not similar to the bimodal CN distribution of typical Milky Way GCs NGC~288 and NGC~362\footnote{See also \cite{milone+17} who classified NGC~362 as type II with a small extra group of 2G stars, difficult to detect here with a sample of 20 stars.}, even though it covers the same range and the stars have similar metallicities. 
We also compare with the CN distribution of the anomalous cluster M~22, that presents two peaks with an uneven distribution. Although it does not have four peaks like NGC~1851, the CN-strong stars are s-rich and CN-weak are s-poor \citep[e.g.][]{marino+11}. The three bottom panels show the KDEs of our sample divided into {\it pec}, {\it S1}, and {\it S2}. We argued before about the possibility of an increasing s-process element abundance from {\it pec} to {\it S2}. Here we show that, in fact, {\it pec} stars are all CN-weak and {\it S2} stars are predominantly CN-strong. Stars of type {\it S1}  cover the full range.

We also checked the stability of the $\delta$S(3839) distribution shape considering another definition for this index applied by \cite{smith+82} that uses slightly different wavelength limits. They analysed the NGC~3201 CN distribution for RGB stars brighter than V $<$ 14 mag and only had one peculiar star in their sample. They found a bimodal distribution with no sharp separation between the peaks. We applied the same strategy as above and scaled S(3839) to $\delta$S(3839) from \cite{smith+82} in the same way we did for our data and for Kayser et al. (Appendix \ref{appK08}).
The resulting curves are shown in Fig. \ref{cnSN82} assuming a constant error bar of 0.05 dex as the KDE bandwidth.  We reproduce the four bottom panels of Fig. \ref{dcn} on the four bottom panels of Fig. \ref{cnSN82}, but now using the same S(3839) definition as in Smith \& Norris for a fair comparison. Their KDE presents three peaks while ours have five. This difference is probably because we have less stars and smaller error bars, therefore our results are more sensitive to individual isolated points. Nevertheless both samples cover the same CN range and have a dominant 1G population and a multi-peak 2G population. If our curves for NGC~3201 are compared between Figs. \ref{dcn} and \ref{cnSN82}, both present a dominant 1G peak and a 2G multi-peak population. The separation in {\it pec.}, {\it S1}, and {\it S2} is also similar.

\cite{norris+83} observed 100 stars in M~22 and we show their CN distribution in Fig. \ref{cnSN82}. It is comparable to the distribution shown in Fig. \ref{dcn} with data from \cite{kayser+08}. The main difference is that Norris \& Freeman have many more stars and their data reveal two additional extreme peaks with CN-stronger and CN-weaker stars.
In other words, M~22 also has a quadrimodal CN distribution with a division of s-rich and s-poor stars as in the case of NGC~1851.
In the same Fig. \ref{cnSN82} we also show the results of NGC~1851 analysed by Campbell et al. and scaled in the same way as in Fig. \ref{fig:tefflogg} for consistency.
The quadrimodal distribution found by \cite{campbell+12} is still present in this plot if the right peak is ignored because it has only one star, and if the two central peaks are considered as one peak. In fact, the peaks found for NGC~1851 in the original paper do not necessarily follow a Gaussian shape. This complexity resembles that of NGC~3201 from our data.

The conclusion is that the shape of the distribution is sensitive to the definition of the index S(3839), sample size, and uncertainties. 
Not surprisingly, the clearly bimodal distribution of NGC~362 from Fig. \ref{dcn} does not look as smooth in \cite{smith83} and \cite{norris87}. Nevertheless, both concluded that this cluster has a bimodal CN distribution. The complexity may come from the fact that photometric indices were used instead of spectroscopic ones.
More recently, \cite{milone+17} have shown that the pair of clusters NGC~288 and NGC~362 do have a very similar chromosome map, but the latter reveals a small population of redder RGB stars, which made this cluster a type-II GC according to their classification. In other words, the CN distribution alone cannot be used to split groups of clusters, but it is certainly a valid piece of information in the vast parameter space that is required to disentangle the multiple populations within a GC and eventually correlate to global properties and environmental effects. In fact, the majority of the clusters analysed by \cite{norris87} with CN distributions, by \cite{carretta+09gir} with Na-O anti-correlations, and by \cite{milone+17} with photometric chromosome maps show two main populations of stars. Any cluster that differs from that behaviour requires special treatment. We have shown in Figs. \ref{dcn} and \ref{cnSN82} that the clusters M~22, NGC~1851, and NGC~3201 do present an odd CN distribution.
S-process element abundances, such as Y, Zr, Ba, La, and Nd in contrast with r-process Eu, and also C+N+O abundances are needed in order to split these sub-populations.

%

\section{Galactic or extragalactic?}
\label{sec:extragal}

Anomalous clusters like M~22, NGC~1851, M~2, and NGC~5286 all present Fe/s-rich and Fe/s-poor stars, even though the Fe-spread is still under discussion for some of them. In all cases where abundances are available, s-rich and s-poor stars also present different abundances of C+N+O. These groups correlate well with an SGB and RGB split if appropriate colours are used in the CMD. Each group has its own Na-O and C-N anti-correlations that are typical signatures of GCs. While the explanation for the anti-correlations seems to be related to self-pollution of second generation stars by the primordial population, the split into Fe/s-rich and Fe/s-poor stars for a few anomalous clusters is not explained by the same mechanisms. \cite{bekki+12} proposed that such clusters could be the result of a merger of two clusters. This is likely to happen in the nucleus of dwarf galaxies where relative velocities and the volume are smaller than that of the Milky Way halo. Should this scenario be true, then the anomalous clusters would have an extragalactic origin from dwarf galaxies captured by the Milky Way (but see also \citealp{marino+15,dacosta15,bastian+18}). We use the expression `extragalactic chemical tagging' to refer to this small group of anomalous clusters, although the accuracy of this term is still debatable and is part of a complex topic of study that has been developed in the last decade.
In this context, we try to tag NGC~3201 as a anomalous GC belonging to this group or not.

In the previous section we showed that NGC~3201 presents three groups on the CN-CH correlation plot, namely {\it pec}, {\it S1}, and {\it S2}, split by 7$\sigma$ in CH. An indication that these groups have different abundances of C+N+O and s-process elements was revealed by photometric indices. Another mild constraint is the CN distribution that seems to be multi-modal for anomalous clusters and this is also the case for NGC~3201. These three points indicate that NGC~3201 may be included in this selected group of globular clusters, with the note that it is possibly a transient object between typical and anomalous GCs, because the features listed above are not extreme for NGC~3201. In fact, even the star-to-star Fe spread has been discussed in the literature \citep[see][]{simmerer+13,covey+03,munoz+13,mucciarelli+15a}. We further note that \cite{dacosta15} included M~22 to this group because it has a spread in s-process element abundances, in spite of its uncertain Fe-spread. NGC~3201 provokes a similar discussion on the Fe-spread but Da Costa did not consider this cluster anomalous because no detailed study on C+N+O and s-process elements was available yet. We have presented some indication from photometry in this direction to be confirmed by high-resolution spectroscopy.

\subsection{Stellar nucleosynthesis {\rm versus} environmental effects}

Environmental effects such as GC merger and extragalactic origin are not the only possible explanation for the more complex GCs with unusual multiple populations. One example is \object{NGC~2808} that presents three to five populations as shown by different analyses \citep[e.g.][and others]{carretta+15,marino+17,milone+17}. Stars in NGC~2808  all have the same metallicity as opposed to M~22 and NGC~1851. The five populations of NGC~2808 are characterised by large variations in light-element abundances, such as Na, O, Mg, and Al. Moreover, it was found that He abundance variations plays an important role in characterising NGC~2808 populations \citep[e.g.][]{bragaglia+10,marino+17}. For reference, we note that \object{NGC~6121} (\object{M~4}) has a similar metallicity of [Fe/H]$\approx$-1.1, but shows a bimodal distribution of Na and CN \citep{marino+08}.

A comparison cluster for NGC~3201 with similar metallicity is NGC~6752, which presents a bimodal CN distribution \citep{norris+81N6752}. However, high-resolution multi-band photometry and spectroscopy revealed a third population for NGC~6752 \citep[e.g.][]{milone+13,nardiello+15,gruyters+14}. The explanation is that the three populations have different light-element abundance ratios and also helium abundances \citep{milone+13,nardiello+15}. In fact, NGC~6752 has a complex horizontal branch morphology that may be related to He abundances \citep{momany+02}. Therefore He, and proton-capture element abundance variations could be enough to explain multiple populations, however heavier elements such as Fe and neutron-capture (and the sum C+N+O) tend to vary within a GC only for anomalous cases such as M~22 and NGC~1851  discussed above. \cite{yong+13} found correlations of s-process elements with Na (that correlates with CN) for NGC~6752, although very small variations of s-process element abundances were measured. No bimodality or clear separation between s-rich and s-poor stars was reported (as is the case for M22 and NGC1851). Yong et al. also speculated that this should be the case for all GCs if they underwent extremely accurate spectroscopic analysis. \cite{yong+15} found no spread on the C+N+O for NGC~6752. In summary, NGC~6752 does not have two populations of s-rich (high C+N+O) and s-poor (low C+N+O), as is the case for M22 and NGC1851 \citep[see also][]{carretta+05}. The three populations found for NGC~6752 with photometry and spectroscopy are related to light-element and helium abundances. NGC~3201 has a similar metallicity and mass to NGC~6752 but differs in many other aspects: (i) it does not show a bimodal CN distribution, but a complex KDE resembling NGC~1851; (ii) CN-CH anti-correlation reveal two trends similar to the findings for M~22, as well as a group of peculiar stars; (iii) it has a much redder horizontal branch (HB) morphology that covers red and blue colours similar to but not as separated as the HB of NGC~1851\footnote{Snapshots of CMDs can be retrieved from here: \url{http://groups.dfa.unipd.it/ESPG/ground.html}.} \citep{mackey+05}. The difference in HB index is consistent with NGC~3201 being 1~Gyr younger than NGC~6752 \citep{rey+01,vandenberg+13,dias+16}.


In conclusion, if the three groups we found for NGC~3201 on our CN-CH analysis have different s-process-element and C+N+O abundances, we may consider it a study case together with the other anomalous GCs in view of the formation scenario of the merger of GCs in the nuclei of dwarf galaxies. If it only possesses a very small variation of s-process-element and C+N+O abundances, and a possible mild trend or correlation of Na(CN) with s-process-element abundances (as for NGC~6752; \citealp{yong+15}), then it is either a typical GC with an odd CN-CH relation or an anomalous GC without C+N+O and s-element abundance variations.
Another possibility is that NGC~3201 could be a transition cluster in the sequence of CN-CH anti-correlation to CN-CH correlation shown by \cite{lim+17}.

High-resolution spectroscopic analysis is needed at least to measure s-process-element and C+N+O abundances of the 28 NGC~3201 stars analysed here.
Another analysis that may help is a second-parameter analysis between NGC~6752 and NGC~3201 to check whether only metallicity and age (plus the complex He contents of the former) are enough to explain the HB morphology differences between the clusters or if NGC~3201 needs to have He abundance variations to produce such HB.

\subsection{Anomalous clusters from the nucleus of dwarf galaxies}

Anomalous clusters are among the most luminous in the Milky Way. Those with M$_{\rm V} \lesssim -7.9$ (i.e. M $\gtrsim 2\times 10^5$ M$_{\sun}$) account for 41 out of 157 from the catalogue of \citet[][2010 edition]{harris96}.
\cite{dacosta15} said that 13 of those are unstudied or poorly studied. In other words, if nine anomalous massive clusters were found among the 28 well-studied massive clusters, we can extrapolate this ratio of 32\% to the total sample from the Harris catalogue and predict that about 13 massive clusters should be anomalous in the Milky Way. This is already about 50\% more than the eight clusters predicted by Da Costa. Instead, if we consider clusters of all masses, the first approximation is to assume that the 70\% of the catalogued clusters that have metallicity measured spectroscopically \citep{dias+16msgr} also have detailed enough studies to detect Fe-spread. In this case 8\% of all clusters should be anomalous, which leads us again to a total of about 13 clusters. Assuming that for half of these clusters there was minimal spectroscopic information, then the number of anomalous GC could be doubled to 26 of all masses. Some effort has been done to find more candidates to host Fe-spread \citep[e.g.][]{saviane+12}. We are still far from having a complete view and a explanation for these peculiar clusters \citep{bastian+18}.

\begin{figure}[!htb]
\centering
\includegraphics[width=\columnwidth,angle=-90]{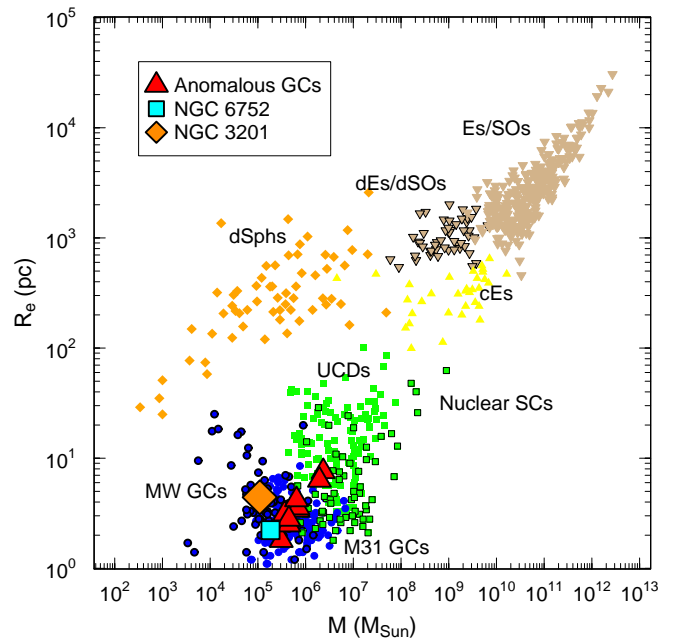}
\caption{Mass-size distribution of compact stellar systems with data from \cite{norris+14}. We highlight the nine anomalous GCs listed by \cite{dacosta15} and \cite{marino+15}. We also feature NGC~3201 and the 
massive GC with no dispersion in Fe or s-process elements, NGC~6752.}
\label{css}
\end{figure}

We compare the structural parameters of NGC~3201 with the anomalous
and to the non-peculiar massive GC NGC~6752 in Fig. \ref{css}. Mass and size were taken from \cite{norris+14}, and when not available for the GC we interpolated from the fitted relation between M$_{\rm V}$ from \citet[][edition 2010]{harris96} and the masses derived by Norris et al. 

\begin{equation}
M_*/M_{\odot} = e^{4.33 -1.008 \cdot M_V} .
\label{mvmo}
\end{equation}

\noindent We note that all anomalous clusters follow a tight mass-size relation in a
region 
on the plot where objects could
be the result of a disruption of a nucleated dwarf galaxy \citep[e.g.][]{bekki+03,pfeffer+13}. As we discussed before, not all massive clusters have the same characteristics and NGC~6752 is plotted to illustrate that not all GCs following this trend are anomalous. 
Surprisingly, NGC~3201 does not belong to the trend of anomalous clusters, despite the other common characteristics discussed in this paper. Should this trend of anomalous clusters in Fig. \ref{css} be the locus of anomalous clusters with a possible extragalactic origin, it could mean that NGC~3201 is the first anomalous GC with a different formation scenario, or that the explanation for the origin of all anomalous clusters should be revised, including NGC~3201.

\subsection{Anomalous clusters from the merger of two clusters}

The small group of CN-weak and CH-weak peculiar stars we found in NGC~3201 is intriguing.
We evaluate here a possible external origin for them.
One possibility is to look at the G4300 indices of NGC~3201 in Fig. \ref{fig:tefflogg} and compare them with those from NGC~362, which has a similar metallicity, in Fig. \ref{n362}. 
Stars from NGC~362  are systematically weaker in CH ($\Delta = 0.1$~mag), with a similar CN. This is basically the difference between {\it pec} and {\it S1}-1G stars in NGC~3201.
We added the peculiar stars of NGC~3201 to the plot of NGC~362 in Fig. \ref{n362} after correcting the magnitudes by the distance modulus difference of 0.63~mag \citep[][2010 edition]{harris96}, for a better visualization. In fact, the peculiar stars are in good agreement with 1G stars of NGC~362; they are CN-weak and CH-strong.
Therefore, the Milky Way has a GC with 1G stars matching the metallicities and CN-CH contents of {\it pec} stars in NGC~3201. Should this scenario of merger be true, there is already a pair of real clusters as constraints for the models.
As a result, CN-CH is useful to find potential mergers.
For Na-O anti-correlation the differences between clusters are less noticeable \citep{carretta+09gir} and Fig. \ref{fig:tefflogg} confirms that peculiar, {\it S1}, and {\it S2} stars all follow similar trends of Na-O.

Another point of view is that there are only 1G peculiar stars and not 2G stars, that 
are typical signatures of GCs. Globular cluster 2G stars are more abundant than 1G stars, so if they were present in the peculiar group, we should have detected them. 
This characteristic is typical in field stars, that are all 1G stars.
In conclusion, we may be looking for an early GC-GC merging of one typical GC with another with only first generation stars, both with the same metallicity or the merging of the progenitor of NGC~3201 with the building blocks of the Milky Way halo (in a single metallicity environment) that almost exclusively presents 1G stars \citep[e.g.][]{fernandez-trincado+16}.
If this scenario is correct, some questions still need to be answered. What is the probability of a merger of two globular clusters with similar metallicity? Is a merger enough to produce a retrograde orbit? Was NGC~3201 formed in the nucleus of a dwarf galaxy? If so, why do we not detect a stellar stream or halo in such an empty region of the sky?

%

\section{Summary and conclusions}

We have obtained low-resolution spectra of RGB stars in the GC NGC~3201, 
derived CN and CH indices, and characterised its multiple stellar generations. 
Three groups were found in the CN-CH relation for the first time for NGC~3201: the main sequence {\it S1}, a secondary less-populated sequence {\it S2}, and a peculiar group {\it pec} with only CN-weak and CH-weak stars. The three groups are separated by 0.1~mag in CH, which is at 7$\sigma$ distance. Although the stars are located around the RGB bump or above, no differences were found in the sequences due to extra mixing,
possibly because NGC~3201 is relatively metal-poor and the convective cell does not reach the H-burning shell to
mix up the CNO-cycle processed material.

Photometric indices were calculated from publicly available catalogues from observations with
ground-based and space-based telescopes to find one that splits the groups {\it pec}-{\it S1}-{\it S2}. 
The key is the use of UV filters that are sensitive to light chemical elements.
The index $\delta$c$_{\rm U,B,I}$ was only able to separate 1G from 2G stars. The V-$\delta$c$_{\rm F275W,F336W,F438W}$ diagram revealed that {\it S1} stars are well split into 1G and 2G along the RGB, and that the smooth transition between the two arms of the RGB havs {\it S2} and {\it pec.} stars, which could explain why this cluster does not show a totally bifurcated RGB. The $\delta$CN-$\delta$c$_{\rm F275W,F336W,F438W}$ correlation was useful to find peculiar stars, and the chromosome map $\delta$c$_{\rm F275W,I}$-$\delta$c$_{\rm F275W,F336W,F438W}$ separates 1G and 2G stars well only for {\it S1}, while {\it S2} and {\it pec} stars are concentrated. Probably the most interesting index is $\delta$c$_{\rm B,V,I}$ that was able to split the groups in this increasing sequence {\it pec}-{\it S1}-{\it S2}. Although s-process elements do not directly affect broad-band magnitudes, for the case of NGC~5286 this index splits s-Fe-rich from s-Fe-poor stars very well. These groups might also correlate with differences in C+N+O, like other anomalous clusters. In fact, it was shown that anomalous clusters have a correlation in CN-CH with increasing abundances of Fe, C+N+O, and s-process elements. High-resolution spectroscopic analysis is needed to confirm this finding for NGC~3201.

Kernel density estimation of CN indices of NGC~3201 stars were compared to other clusters. We note that the distributions are sensitive to the size sample, uncertainties, and definition of the index, therefore the evidence found in this analysis is mild.
We conclude that anomalous GCs like M~22 and NGC~1851 show a quadrimodal CN distribution while typical Milky Way GCs, like \object{NGC~288} and \object{NGC~362,} reveal a bimodal CN distribution. We uncover a quadrimodal 
CN distribution for NGC~3201, resembling that of anomalous clusters.

NGC~3201 cannot be classified as a typical Milky Way GC, but its inclusion in the anomalous group needs confirmation by high-resolution spectroscopic analysis.
 So far, all anomalous clusters present star-to-star variation of Fe, C+N+O, and s-process elements, where each group has its own anti-correlations of p-capture elements (Na-O, CN-CH), whenever these abundances are available. NGC~3201 is an exceptional case with multiple sequences of CN-CH that seems to correlate with C+N+O and s-process-element abundances, but with no Fe-spread (or at least debatable as in the case of M~22). 

A possible scenario for the origin of anomalous clusters is in the nucleus of dwarf galaxies.
Anomalous GCs seem to follow a trend on the mass-size relation. Some models have shown how larger galaxies may evolve to ultra compact dwarfs and how dwarf galaxies can evolve to globular clusters.
NGC~6752 illustrates that not all clusters in this trend are anomalous. NGC~3201 does not follow this trend. If the locus of an anomalous cluster is only in this trend of mass-size, either the scenario needs revision to incorporate NGC~3201, or NGC~3201 is not anomalous but a rather abnormal cluster.

Four peculiar stars were discovered presenting CN-weak and CH-weak indices. One of such stars 
was already indicated by \cite{smith+82} and \cite{dacosta+81}.
The nature and origin of this group of stars needs further study with high-resolution spectroscopy
and detailed chemical abundance analysis. Our proposal scenario to explain these stars is that they
were accreted to NGC~3201 during a GC-GC merger, where one of them had
only 1G stars, or because NGC~3201 was born in the nucleus of a dwarf galaxy and accreted 1G field stars. All stars should have the same metallicity and similar distribution of CN indices, but with a systematic difference in CH indices.

We compared NGC~3201 with NGC~6752 that has similar mass and metallicity. The latter follows the mass-size relation of anomalous clusters but it has its complexities in the horizontal branch (HB) driven by He abundances.
The difference in HB morphology can be explained by the age difference of 1~Gyr.
Despite the similarities in global parameters, NGC~6752 does not show complexities similar to the anomalous GCs. In other words, a combination of mass, size, metallicity, age, and nucleosynthesis does not mean that a GC is anomalous. The origin of the anomalies of NGC~3201 and anomalous GCs seems to be external.

Finally, we envisage three possibilities for NGC~3201:

\begin{enumerate}
\item{CN-CH {\it pec-S1-S2} is an increasing sequence of C+N+O and s-process-element abundances, which means that NGC~3201 would be the first anomalous GC out of the mass-size relation.}

\item{{\it pec-S1-S2} have the same contents of C+N+O and s-process-element abundances, which means that NGC~3201 would be the first non-anomalous GC with multiple CN-CH anti-correlation.}

\item{{\it pec-S1-S2} have the same contents of C+N+O and s-process-element abundances, and NGC~3201 would be the first anomalous GC without star-to-star C+N+O and s-process-element abundances.}

\end{enumerate}

In all cases, the definition of anomalous clusters and the scenario in which they have an extragalactic origin must be revised.

\begin{acknowledgements}
This work was only possible because of ESO approval for the ESO/NEON Observing School at La Silla on February 2016. The authors acknowledge M. Dennefeld and F. Selman for leading the organisation of the school. JPNC thanks E. Pancino for the discussion on the error estimation. The authors are thankful to Prof. G. Da Costa for a careful reading and comments on the manuscript, and also to the anonymous referee for useful suggestions that helped improve the paper. 
\end{acknowledgements}


\bibliographystyle{aa} 
\bibliography{bibliography} 

\appendix
\section{Scaling indices from the literature}
\label{appK08}

\subsection{\cite{kayser+08}}

We transformed the indices S(3839) and G4300 and their uncertainties from \cite{kayser+08} from natural logarithm to common logarithm. Using the new indices we followed the procedure described in Sect. \ref{sec:tefflogg} to correct the indices from stellar surface temperature and gravity effects. The goal is to put all the results in the same scale to allow direct comparisons between our results and those from \cite{kayser+08}.

\begin{figure}[!htb]
\centering
\includegraphics[angle=-90,width=0.8\columnwidth]{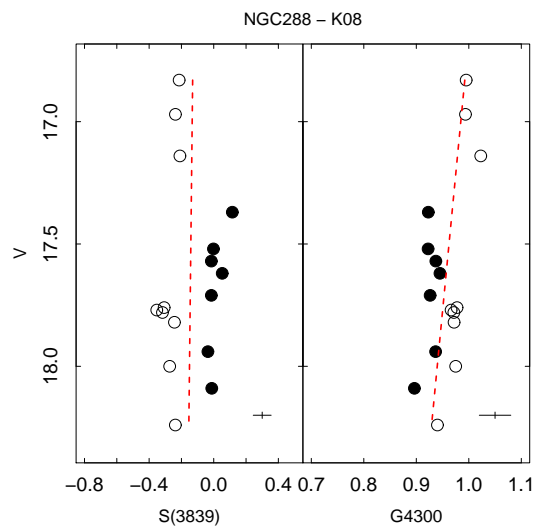}
\caption{Similar to Fig. \ref{fig:tefflogg} but for NGC~288 with parameters from \cite{kayser+08}. Linear fit because bright stars are only CN-weak. Filled circles are CN-strong, and empty circles are CN-weak stars. Average error bars are displayed in the bottom right corner of each panel.}
\label{n288}
\end{figure}

\begin{figure}[!htb]
\centering
\includegraphics[angle=-90,width=0.8\columnwidth]{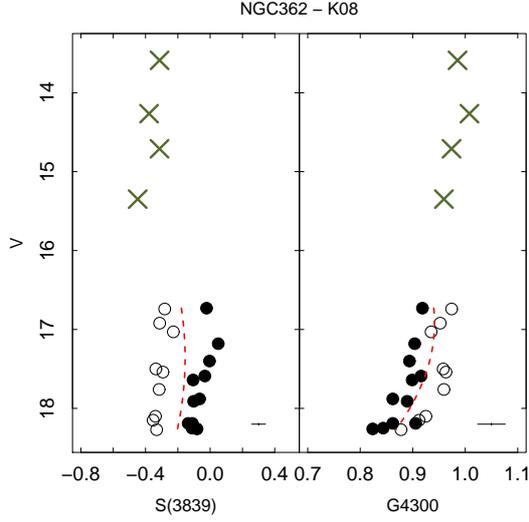}
\caption{Same as Fig. \ref{n288} but for NGC~362. Green crosses are the peculiar stars of NGC~3201 for discussion in Sect. \ref{sec:multipop}. Magnitudes of NGC~3201 stars are shifted by 0.63~mag to match the distance modulus of NGC~362.}
\label{n362}
\end{figure}

\begin{figure}[!htb]
\centering
\includegraphics[angle=-90,width=0.8\columnwidth]{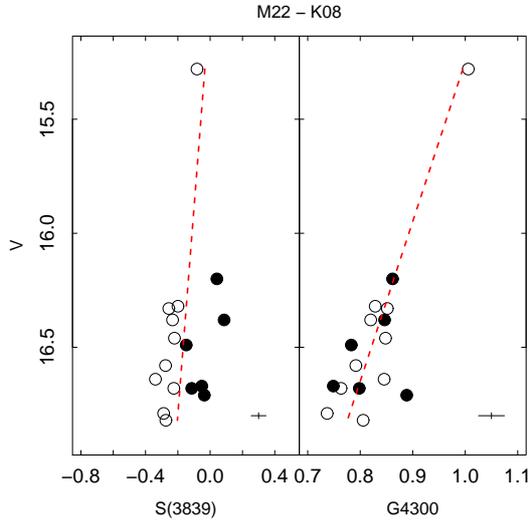}
\caption{Same as Fig. \ref{n288} but for M~22.}
\label{m22}
\end{figure}

\subsection{Indices defined by \cite{norris+81}}

In Section \ref{sec:cndist} we also use indices following the definition of \cite{norris+81}. We follow the same procedure as in Fig. \ref{fig:tefflogg} to put into the same scale the CN indices of NGC~1851 \citep{campbell+12}, M~22 \citep{norris+83}, and NGC~3201 \citep{smith+82}. Figures are shown below.

\begin{figure}[!htb]
\centering
\includegraphics[angle=-90,width=\columnwidth]{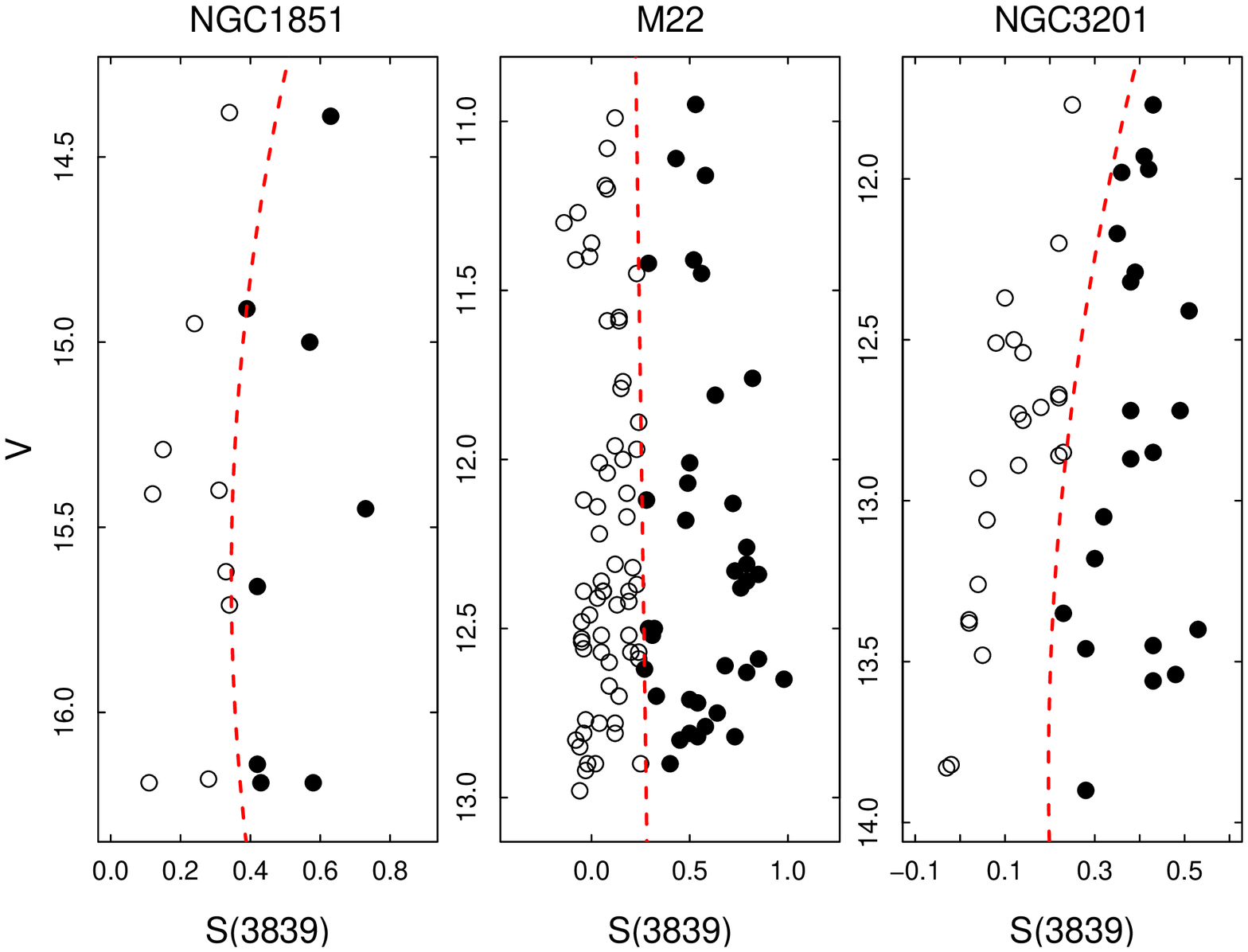}
\caption{Similar to Fig. \ref{fig:tefflogg} but only for CN of NGC~1851 \citep{campbell+12}, M~22 \citep{norris+83}, and NGC~3201 \citep{smith+82}. Filled circles are CN-strong and empty circles are CN-weak stars.}
\label{teffloggSN82}
\end{figure}

\section{Errors of the indices}
\label{appError}

In order to derive the equations for the errors of the indices, we adapted the work realised by \cite{vollmann+06}, who obtained these equations for equivalent widths. The index ($I$) definition for a general case, taking into account one or $n$ continua regions, is expressed by 

\begin{equation}
I = -2.5 \log \left( \frac{\int^{\lambda_{2}}_{\lambda_{1}} F(\lambda) \, d \lambda}{\frac{1}{n} \sum_{i=1}^{n} \int^{\lambda_{2,i}}_{\lambda_{1,i}} F_{c}(\lambda) \, d \lambda} \right)
\end{equation}

\noindent with $F(\lambda)$ the flux in the band and $F_{c}(\lambda)$ the flux in the continuum at the wavelength $\lambda$.  Then, defining $\Delta \lambda = \lambda_{2} - \lambda_{1}$ and $\Delta \lambda_{c_{i}} = \lambda_{2,i} - \lambda_{1,i}$,  and applying the mean value theorem, the  expression can be written as

\begin{equation}
\label{Index}
I = -2.5 \log \left(  \frac{\bar{F} \, \Delta \lambda}{\frac{1}{n} \sum_{i=1}^{n} \bar{F}_{c_{i}}\, \Delta \lambda_{c_{i}}}  \right), 
\end{equation}

\noindent where $\bar{F}$ and $\bar{F}_{c_{i}}$ are the arithmetic mean within $\Delta \lambda$  and $\Delta \lambda_{c_{i}}$ of the band and continua fluxes, respectively. 

Following the principle of error propagation, as \cite{vollmann+06} did,  we expand the last equation in a Taylor series, 

\begin{equation}
I= I(\bar{F}, \bar{F}_{c_{1}}, ..., \bar{F}_{c_{n}}) + \frac{\partial I}{\partial \bar{F}} \left( F - \bar{F} \right) + \sum^{n}_{i=1}  \frac{\partial I}{\partial \bar{F}_{c_{i}}} \left( F_{c_{i}} - \bar{F}_{c_{i}} \right), 
\end{equation}

\noindent where $F$ and $F_{c_{i}}$ are random variables. The variance of the expansion is 

\begin{equation}
\label{error}
\sigma^{2}(I)= \left( \frac{\partial I}{\partial \bar{F}} \cdot \sigma(F) \right)^{2} + \sum^{n}_{i=1} \left(   \frac{\partial I}{\partial \bar{F}_{c_{i}}} \cdot \sigma(F_{c_{i}})  \right)^{2},
\end{equation}

\noindent with $\sigma(F)$ and $\sigma(F_{c_{i}})$ the standard deviation in the band and continua, respectively. If we assume a Poisson statistic, the standard deviations are defined by

\begin{equation}
\sigma(F_{c_{i}}) = \frac{\bar{F}_{c_{i}}}{\mathrm{S/N}}
\end{equation}

\noindent and

\begin{equation}
\sigma(F)= \sqrt{\frac{\bar{F}}{\bar{F}_{c_{i}}}} \cdot \sigma(\bar{F}_{c_{i}}) = \frac{ \sqrt{\bar{F}\, \bar{F}_{\bar{c}}}}{\mathrm{S/N}},
\end{equation}

\noindent where $\bar{F}_{\bar{c}}$ is the arithmetic mean of every $\bar{F_{c_{i}}}$. The partial derivatives, using Equation \ref{Index}, are 

\begin{equation}
\frac{\partial I}{\partial \bar{F}} = \frac{-2.5}{\bar{F} \ln (10)}
\end{equation}

\noindent and

\begin{equation}
\frac{\partial I}{\partial \bar{F}_{c_{i}}} = \frac{2.5}{\ln (10)} \frac{\Delta \lambda_{c_{i}}}{\sum^{n}_{j=1} \bar{F}_{c_{j}} \, \Delta \lambda_{c_{j}}}.
\end{equation}

Finally, we can write Equation \ref{error} as

\begin{equation}
\sigma(I)= \frac{2.5}{\ln(10)\, \mathrm{S/N}} \left[  \frac{\bar{F}_{\bar{c}}}{\bar{F}} + \sum^{n}_{i=1} \left(   \frac{\bar{F_{c_{i}}}\, \Delta\lambda_{c_{i}}}{\sum^{n}_{j=1} \bar{F_{c_{j}}}\, \Delta\lambda_{c_{j}}} \right)^{2}  \right]^{1/2},
\label{finalerrors}
\end{equation}

\noindent which is the error of an index with $n$ continua regions.

\begin{figure}[!htb]
\centering
\includegraphics[angle=-90,width=0.8\columnwidth]{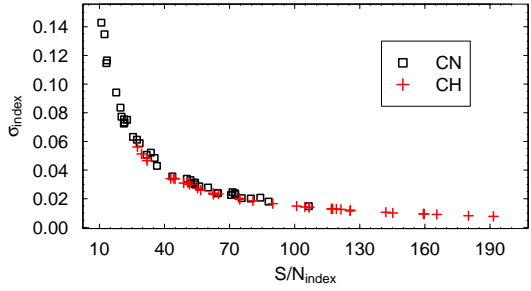}
\caption{Errors from Eq. \ref{finalerrors} as a function of signal-to-noise ratio for CN and CH indices S(3839) and G4300.}
\label{snsig}
\end{figure}

The errors of indices S(3839) and G4300 as a function of signal-to-noise ratio are given in Fig. \ref{snsig}. This relation is useful to plan observations for other objects estimating the minimum S/N to reach the desired precision. In this work we used only spectra with S/N$_{\rm CN} > 25$ (i.e. S/N$_{\rm CH} > 55$) in our analysis.

\end{document}